\documentclass[preprint,12pt,square,comma,number]{elsarticle}
\usepackage{numcompress}
\usepackage{amsmath}
\usepackage{color,txfonts}
\usepackage{amssymb}
\usepackage{graphicx}
\usepackage{amssymb}
\usepackage{verbatim}
\usepackage{booktabs}
\usepackage{subfig}
\usepackage[colorlinks,
            linkcolor=blue,
            anchorcolor=blue,
            citecolor=blue]{hyperref}

\usepackage{lineno}

\usepackage{listings}
\usepackage{xcolor}
\definecolor{mygreen}{rgb}{0,0.6,0}
\definecolor{mygray}{rgb}{0.5,0.5,0.5}
\definecolor{mymauve}{rgb}{0.58,0,0.82}
\begin{document}

\begin{frontmatter}

\title{Simulation of DAMPE silicon microstrip detectors in the $\rm Allpix^{2}$ framework}

\author[1,2]{ Yu-Xin Cui}

\author[1,2]{ Xiang Li}

\author[1]{ Shen Wang  \corref{cor1}}
\ead{wangshen@pmo.ac.cn}

\author[1,2]{ Chuan Yue \corref{cor1}}
\ead{yuechuan@pmo.ac.cn}

\author[1,2]{ Qiang Wan}

\author[1]{ Shi-Jun Lei}

\author[1,2]{ Guan-Wen Yuan}

\author[1]{ Yi-Ming Hu}

\author[1]{ Jia-Ju Wei}

\author[1,2]{ Jian-Hua Guo}
\cortext[cor1]{Corresponding author}

\address[1]{Key Laboratory of Dark Matter and Space Astronomy, Purple Mountain Observatory, Chinese Academy of Sciences, Nanjing 210008, China}
\address[2]{School of Astronomy and Space Science, University of Science and Technology of China, Hefei 230026, China}

\begin{abstract}
Silicon strip detectors have been widely utilized in space experiments for gamma-ray and cosmic-ray detections thanks to their high spatial resolution and stable performance. For a silicon micro-strip detector, the Monte Carlo simulation is recognized as a practical and cost-effective approach to verify the detector performance. In this study, a technique for the simulation of the silicon micro-strip detector with the $\rm Allpix^{2}$ framework is developed. By incorporating the electric field into the particle transport simulation based on Geant4, this framework could precisely emulate the carrier drift in the silicon micro-strip detector. The simulation results are validated using the beam test data as well as the flight data of the DAMPE experiment, which suggests that the $\rm Allpix^{2}$ framework is a powerful tool to obtain the performance of the silicon micro-strip detector.
\end{abstract}


\end{frontmatter}

\section{Introduction}\label{}
With its high spatial resolution and stable performance, the silicon strip detector has been widely used in gamma-ray and cosmic-ray detection experiments in space, such as PAMELA\cite{pamela}, AGILE\cite{agile}, Fermi-LAT\cite{fermi}, AMS-02\cite{ams} and DAMPE\cite{dampe}. For these space instruments, the performance of the silicon strip is essential.

\begin{figure}[!h]
\centering
\subfloat[\label{fig:enegy_com}]{\includegraphics[scale=0.32]{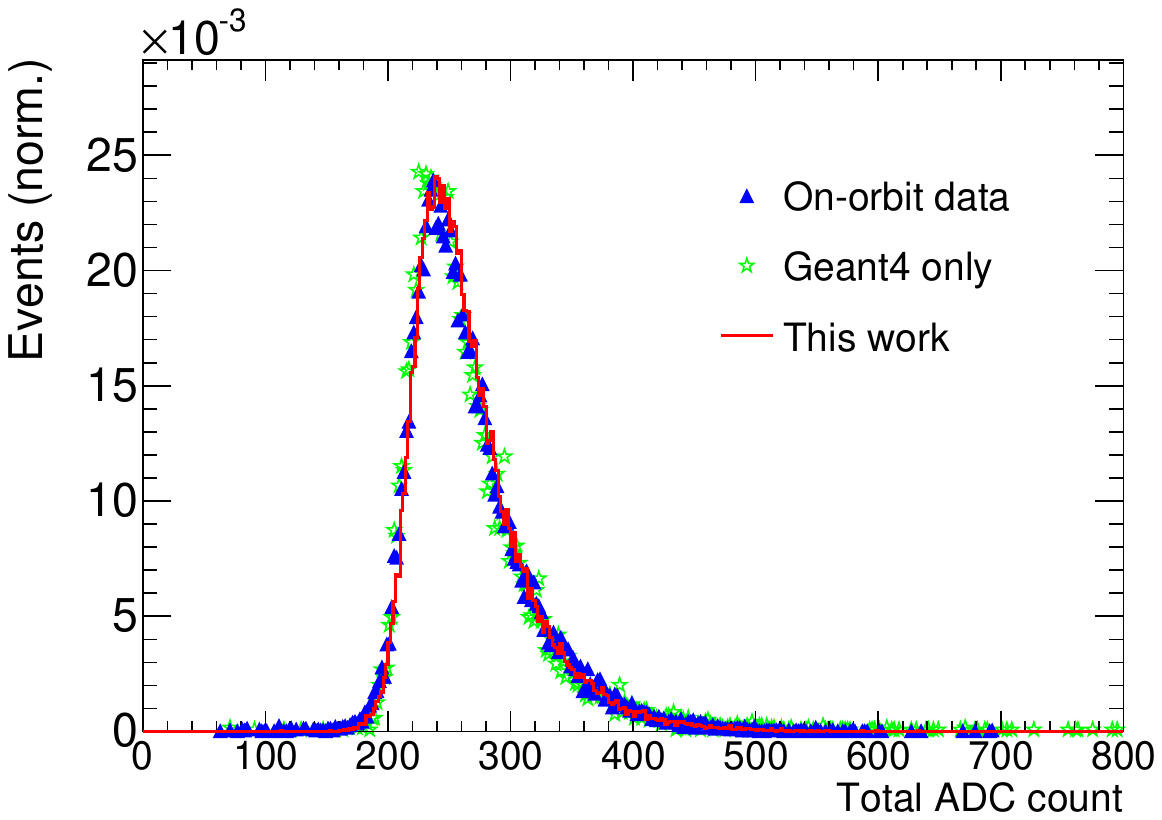}}
\subfloat[\label{fig:num_com}]{\includegraphics[scale=0.32]{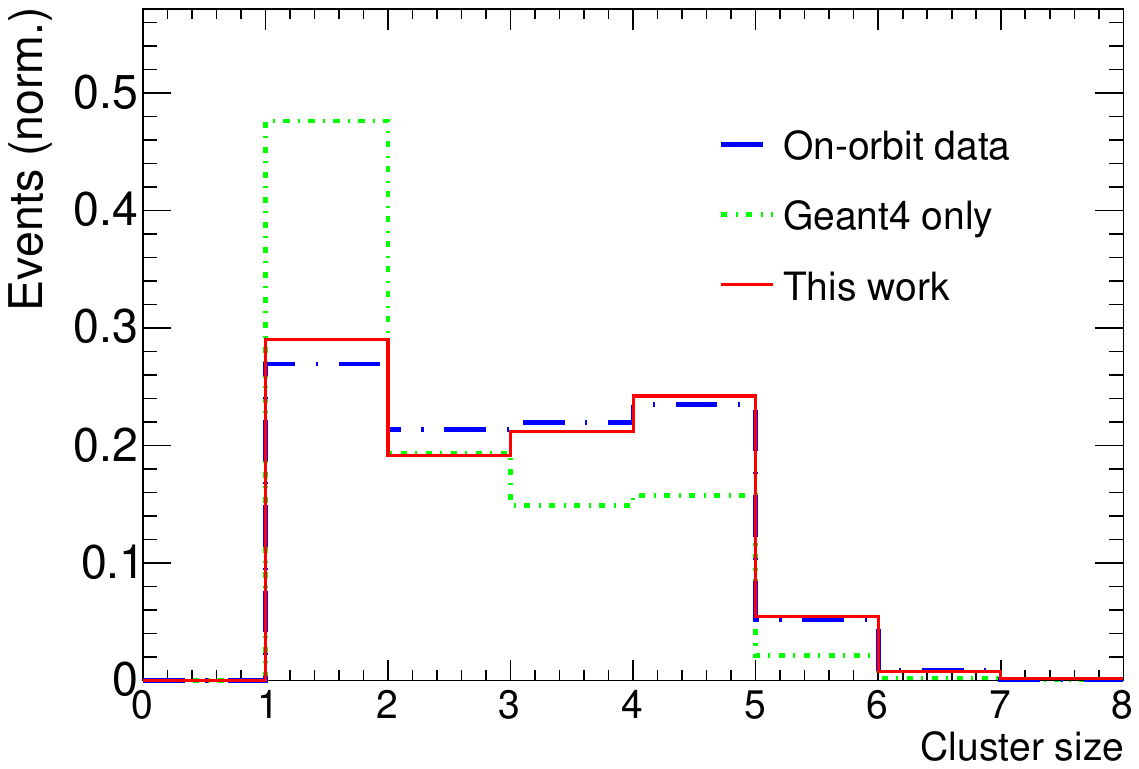}}
\caption{The comparison between on-orbit data and simulated data for vertically incident helium MIP (Minimum Ionizing Particle) entries. The left figure represents the total ADC count of the readout strips, while the right figure depicts the difference in cluster size. The normalization method used in this study is area normalization.}
\label{fig_com}
\end{figure}

Monte Carlo simulation-based verification of the detector performance is a crucial and cost-effective method to predict and verify detector performance. In most cases, using Geant4 to simulate silicon strip detectors is sufficient~\cite{dampe_geant4,cite_simulation}. When comparing simulated data with on-orbit data, Fig.~\ref{fig:enegy_com} shows that the energy deposition of particles in individual strips generally matches well between the simulated and on-orbit data.  However, challenges arise when it comes to charge reconstruction using the charge sharing algorithm in \cite{charge_qiao} due to differences in cluster size (the number of strips firing in an event), as shown in Fig.~\ref{fig:num_com}. These differences can be attributed to the complex internal electric fields present in silicon strip detectors, which are difficult to accurately consider in Geant4 simulations. Even when considering a simple linear electric field, there is still a significant discrepancy between the simulated results and the data \cite{allpix_sim}. The limitations of Geant4 in capturing the intricate behavior of silicon strip detectors necessitate alternative approaches.

One alternative approach is to employ the TCAD software in simulation and modeling. TCAD takes into account the influence of electric fields and provides more accurate results, making it valuable for radiation damage studies and new detector designs \cite{TCAD_damage,TCAD_design}. However, TCAD has limitations in simulating the large number of readout strips present in detectors like DAMPE (which consists of 73,728 readout strips). Integrating TCAD simulations with Geant4 to assess the overall performance, including geometric acceptance and trigger efficiency of the detector, becomes challenging due to these limitations. To address these challenges, we require software capable of considering the effects of internal electric fields, doping concentration, implantation depth, pitch width, and other material properties of the silicon strip detector. Additionally, the software should facilitate large-scale simulations incorporating various detector materials. In this context, the highly successful application of the $\rm Allpix^{2}$ \cite{allpix} framework becomes relevant \cite{allpix_apply,allpix_apply3}.

$\rm Allpix^{2}$ is a user-friendly software package designed to simulate semiconductor detector performance~\cite{allpix}. It covers the entire simulation chain, from ionizing radiation interaction with the sensor to the digitization of hits in the readout chip \cite{allpix_web}. In our work, we utilized the $\rm Allpix^{2}$ framework to simulate the silicon strip detector of DAMPE, which offers several advantages. The increased precision of simulation results enables improved charge and vertex reconstruction for fragmented events, leading to a deeper understanding of the underlying physics processes within the detector. Accurate simulation results can improve background estimation and facilitate the identification of fancy events with DAMPE. In our future research, we plan to utilize $\rm Allpix^{2}$ to investigate radiation damage and charge collection efficiency. Moreover, for future satellite missions like VLAST \cite{vlast}, TCAD simulations can be integrated into the $\rm Allpix^{2}$ framework to evaluate the overall performance of detectors with different silicon strip structures, aiding in the design of detectors that align with specific scientific objectives.

In the subsequent sections, we will provide a comprehensive overview of the design of silicon strip detectors and outline the simulation workflow using the $\rm Allpix^{2}$ framework.

\section{Detector configuration}\label{sec.stk_cfg}
\begin{figure}[!h]
\centering
\includegraphics[scale=0.43]{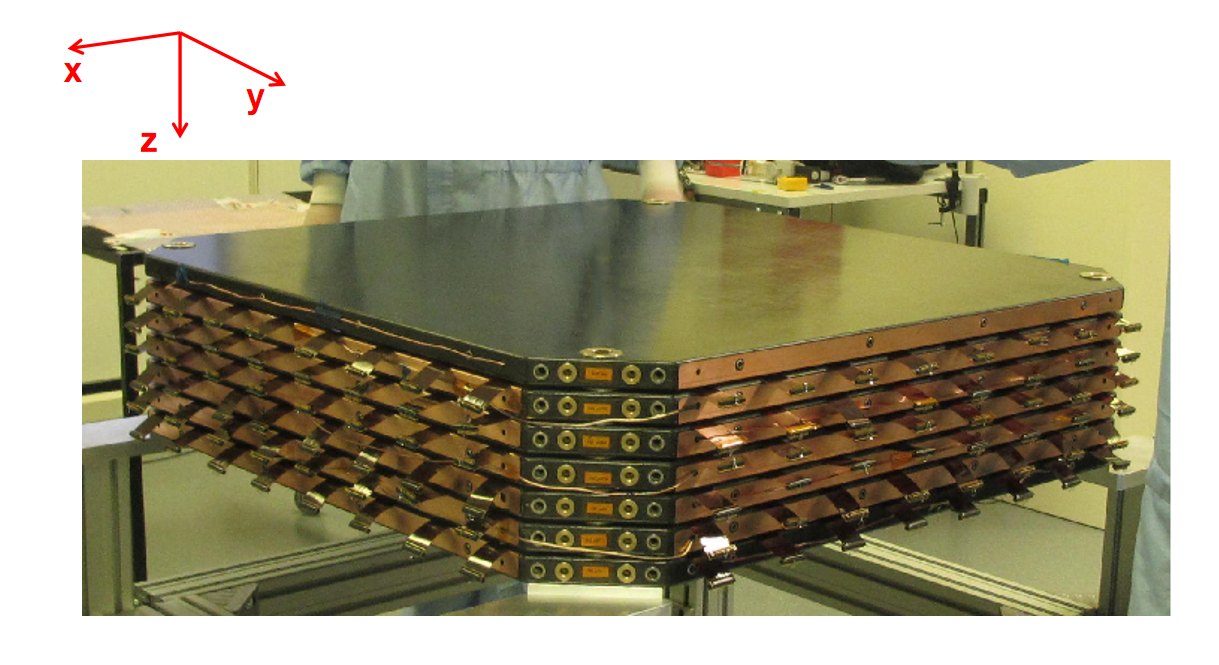}
\caption{A schematic of the STK.}
\label{STK}
\end{figure}

The Silicon Tungsten tracKer (STK), as shown in Fig.~\ref{STK}, is a sub-detector of DAMPE \cite{dampe_stk} to measure the track and charge of an incident particle. It is made up of six planes, each consisting of two orthogonal layers of single-sided silicon micro-strip detectors measuring the hit point in both XZ and YZ dimensions. Three layers of 1 mm-thick tungsten plates are placed in the top three silicon planes for the photon conversion.

\begin{figure}[!h]
\centering
\includegraphics[scale=0.32]{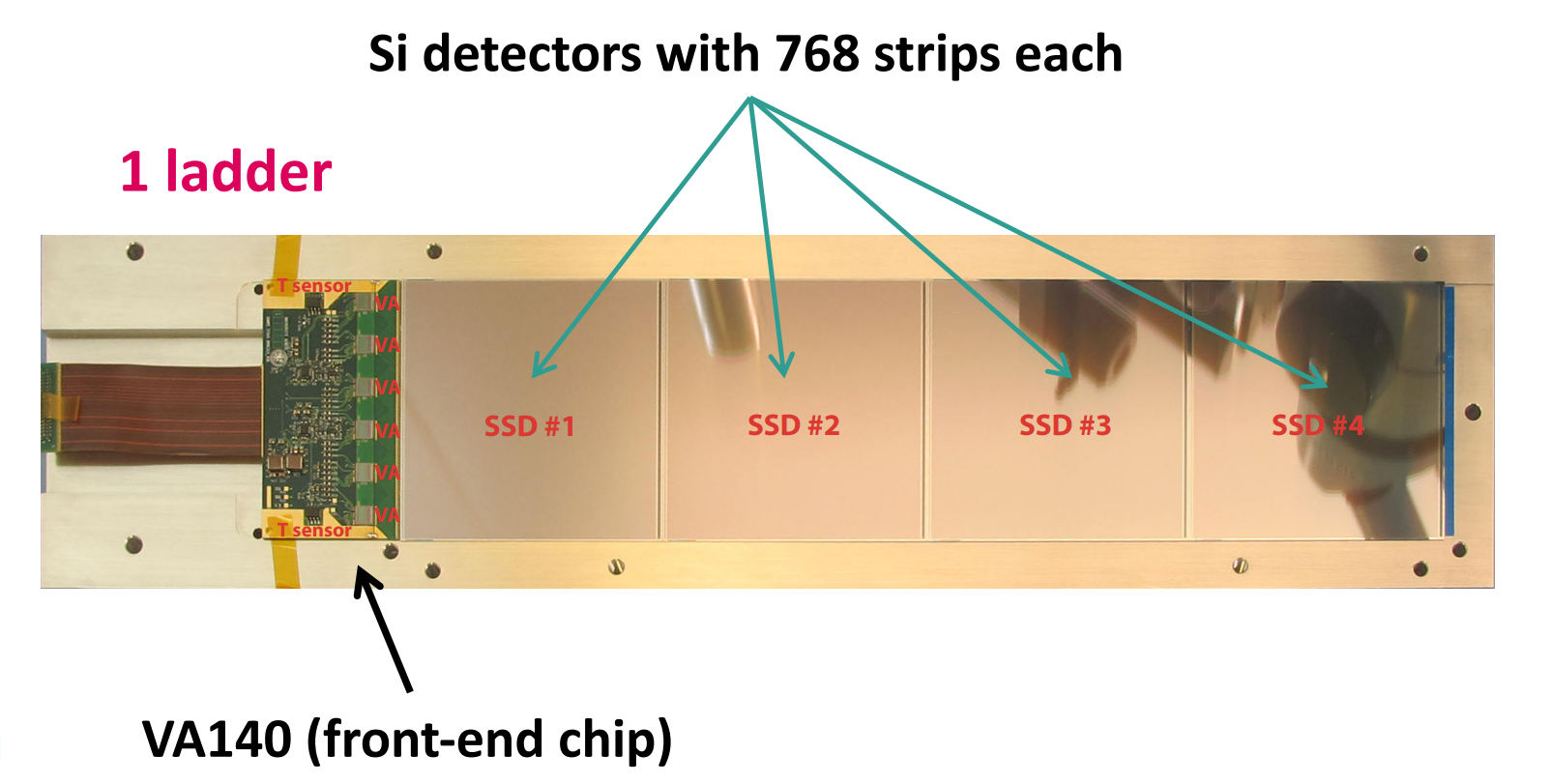}
\caption{A schematic of the ladder.}
\label{ladder}
\end{figure}

In each layer of the STK, there are 16 ladders, each consisting of four Silicon Strip Detectors (SSDs), as shown in Fig.~\ref{ladder}. Each SSD has dimensions of $95\times 95\times0.32$ $mm^{3}$ and is divided into 768 strips with a pitch of 121 $\mu m$. The front-end hybrid (TFH) incorporates six ASIC (Application Specific Integrated Circuit) chips, specifically the VA140 chips manufactured by IDEAS \cite{ideas}, which are responsible for signal shaping and amplification. To optimize the readout process, every other strip is read out, reducing the number of readout channels while preserving a high level of spatial resolution. 

\section{Simulation progress}\label{sec.sim}
\subsection{Geometry and Geant4 configuration}
\begin{figure}[!h]
\centering
\includegraphics[scale=0.51]{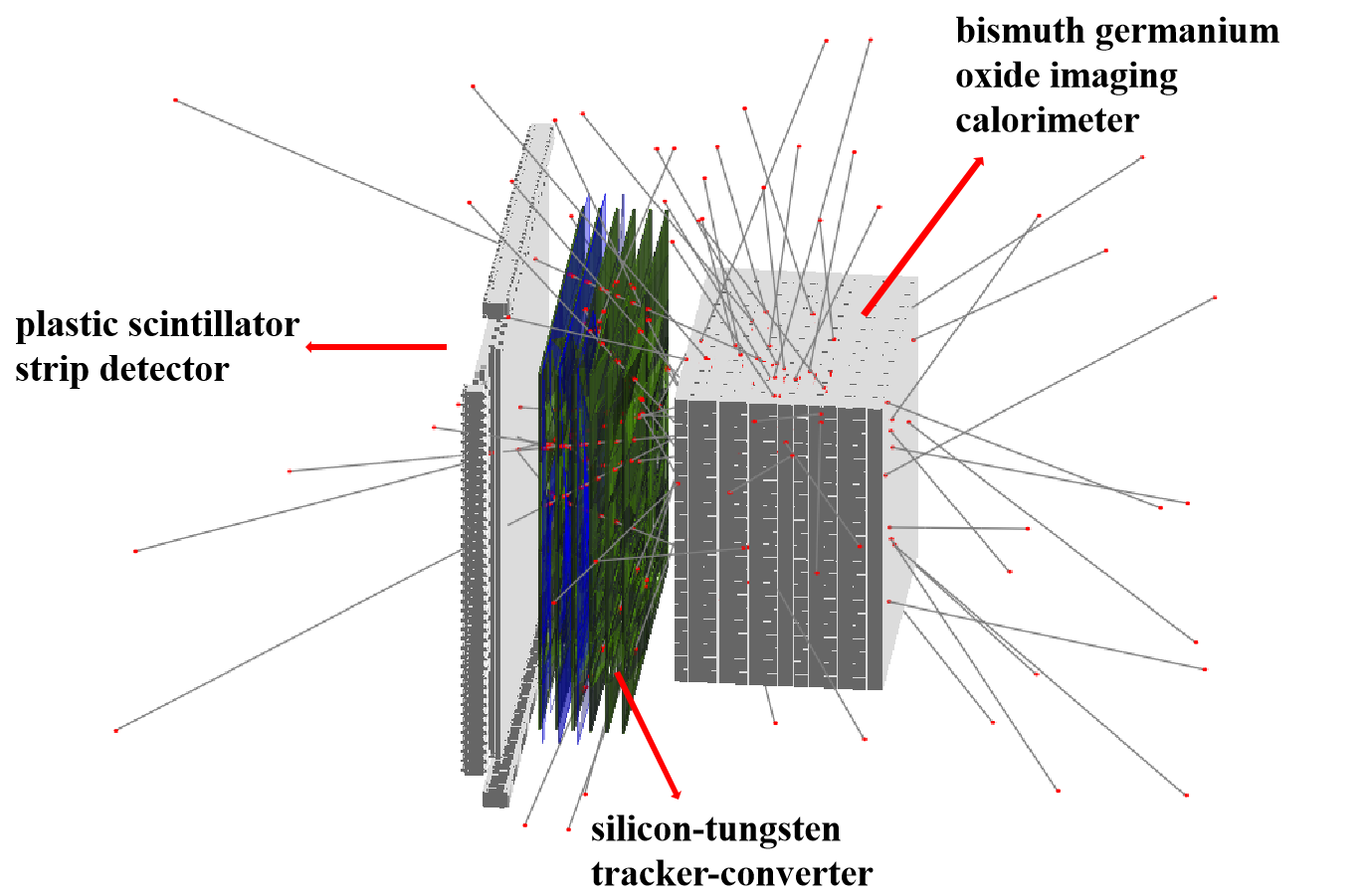}
\caption{The detector geometry with some generated proton events.}
\label{geo}
\end{figure}

The geometry of the simulated STK is constructed in detail according to the configuration of DAMPE. The GDML files of the other two important sub-detectors of DAMPE, the PSD (plastic scintillator strip detector), and the BGO (bismuth germanium oxide) imaging calorimeter, are also imported to achieve good simulation performance. The detector geometry, together with some generated proton events, is shown in Fig.~\ref{geo}. The detector readout modules are set to a ladder configuration in the simulation, which has the same size and position as the STK. The ladder has a length of 380 mm and a width of 768$\times$121 $\mu$m, with 200 $\mu$m of PCB material at the bottom.

Proton and helium MIP events in on-orbit data are selected as the two most abundant particles in cosmic-rays to ensure sufficient statistics for the validation analysis. These two types of particles are generated in the Geant4 simulation. The input energy spectra of proton and helium are obtained from measurements conducted by DAMPE and AMS-02 \cite{He_mips}.

The particle source is a spherical source centered at the origin of the detector coordinate system with a radius of 1 meter. The angular distribution of the particles is isotropic. Geant4 utilizes the FTFP\_BERT\_EMZ physics list for the simulation. In $\rm Allpix^{2}$, the software generates the corresponding charge carriers based on the energy deposited by the particles. The number of electron-hole pairs produced is determined by the material properties of the detector itself. Fluctuations in the charge production are modeled using a Fano factor, assuming Gaussian statistics \cite{allpix_geant4_web}.

\subsection{TCAD Electric field}\label{field}
The Geant4 toolkit is not equipped to handle charge drift in an external electric field, thereby preventing a precise simulation of the charge distribution in each readout strip. To overcome this challenge, we applied the $\rm Allpix^{2}$ simulation framework to simulate the charge drift process \cite{allpix_sim} by importing a user-defined electric field.

\begin{figure}[!ht]
\centering
\includegraphics[scale=0.34]{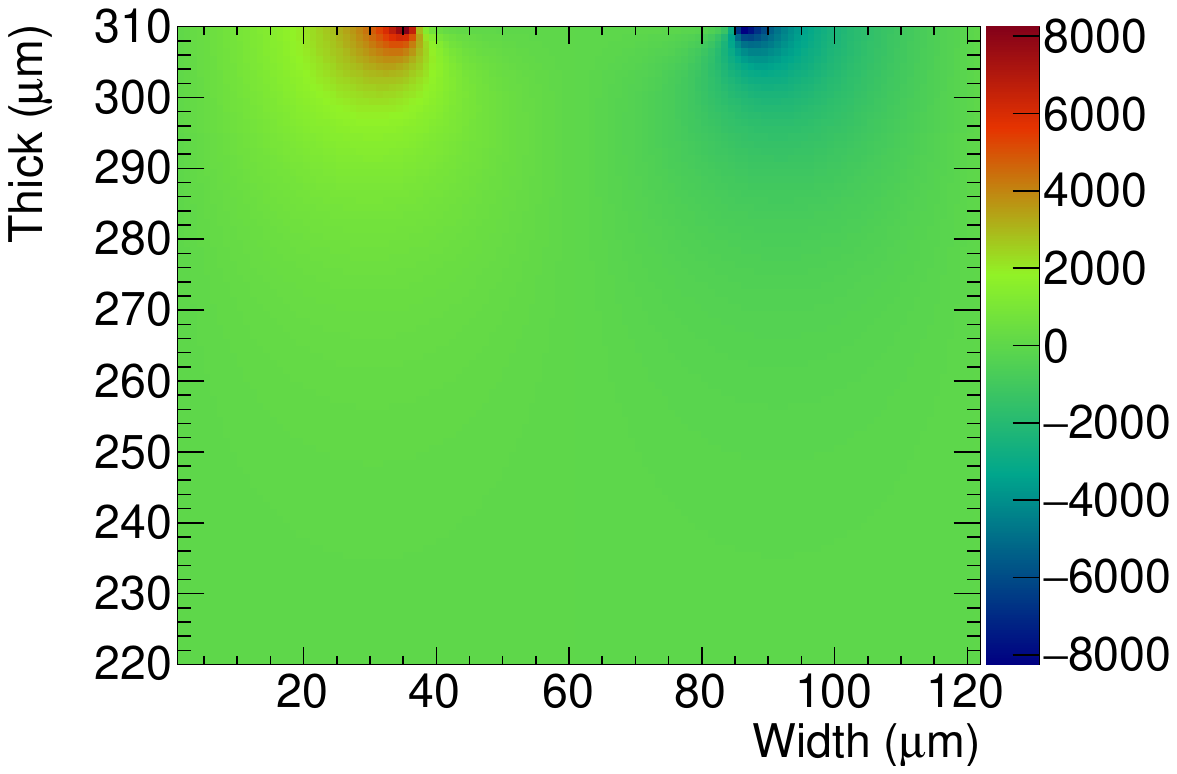}
\includegraphics[scale=0.34]{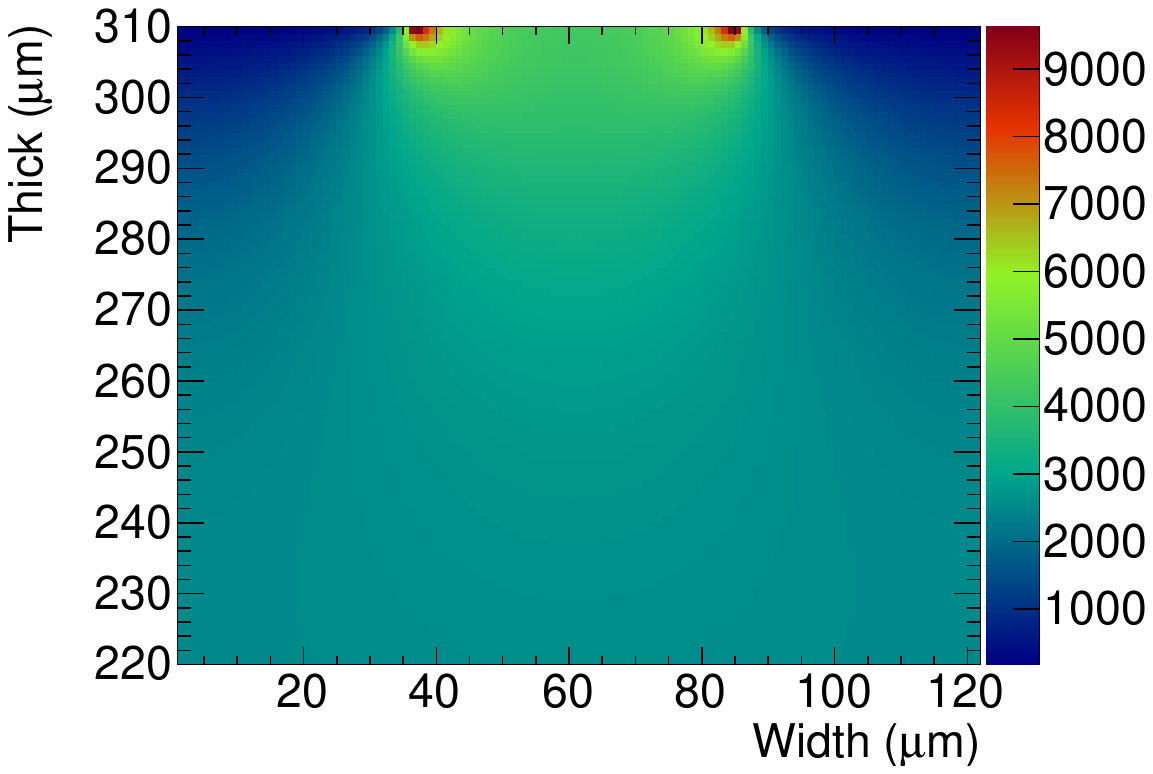}
\caption{The electric field distribution of the TCAD simulation is shown in the X and Y directions, respectively. Only the range of 90 um which exhibits the greatest change in electric field is presented here. The unit of electric field strength is V/m.}
\label{electric field}
\end{figure}

In this work, the electric fields of three pitches are obtained via the TCAD software based on the sensor structure. Then, the electric field of the middle pitch is selected and converted into regular grids using an interpolation algorithm ($\rm Allpix^{2}$ provides tools for converting different types of electric field files into the required format). Finally, the gridded electric field is imported into $\rm Allpix^{2}$ for the charge transport simulation. The output electric field distribution from the TCAD software is illustrated in Figure \ref{electric field}.

\subsection{Charge carrier transport}
The propagation consists of a combination of drift and diffusion simulations. The drift is calculated using the charge carrier velocity derived from the charge carrier mobility and the magnetic field through the calculation of the Lorentz drift. The diffusion is accounted for by applying an offset drawn from a Gaussian distribution, which is calculated based on the Einstein relation:
\begin{equation}
\sigma=\sqrt{\frac{2 k_b T}{e} \mu t}
\end{equation}
using the carrier mobility $\mu$, the temperature $T$ and the time step $t$. The propagation stops when the set of charges reaches any surface of the sensor. The main parameters involved in this process include the following:
\begin{itemize}
\item \emph{temperature} \\
This parameter is set to 277 K to match data from the temperature sensor equipped on the STK of DAMPE.

\item \emph{mobility\_model} \\
$\rm Allpix^{2}$ implements a variety of charge carrier mobility models, including the Jacoboni-Canali Model\cite{module_jac}, the Canali Model\cite{module_can}, the Hamburg Model\cite{module_ham} and others. The software also provides the option to use fully custom mobility models.  The user would define the carrier mobility functions according to the local electric field and the local doping concentration. The best-suited model depends on the simulated device and other simulation parameters. The default Jacoboni-Canali model is used here.

\item \emph{charge\_per\_step} \\
This parameter defines the maximum number of electrons for which the randomized diffusion is calculated collectively. It allows propagation of sets of charge carriers together in order to speed up the simulation while maintaining the required accuracy.

\item \emph{integration\_time} \\
The integration time of the VA140 is 6.5 $\mu s$, during which the generated carriers are collected. During the simulation, this parameter is set to 80 ns, which is enough time for all carriers to reach the electrode.

\item \emph{propagate\_electrons, propagate\_holes} \\
The resultant electrons will drift towards the cathode depending on the applied voltage, whereas the electrode receives a signal generated by electron and hole pairs, requiring both  "propagate\_electrons" and "propagate\_holes" to be set to "true".
\end{itemize}

\begin{figure}[!ht]
\centering
\subfloat{\includegraphics[scale=0.1]{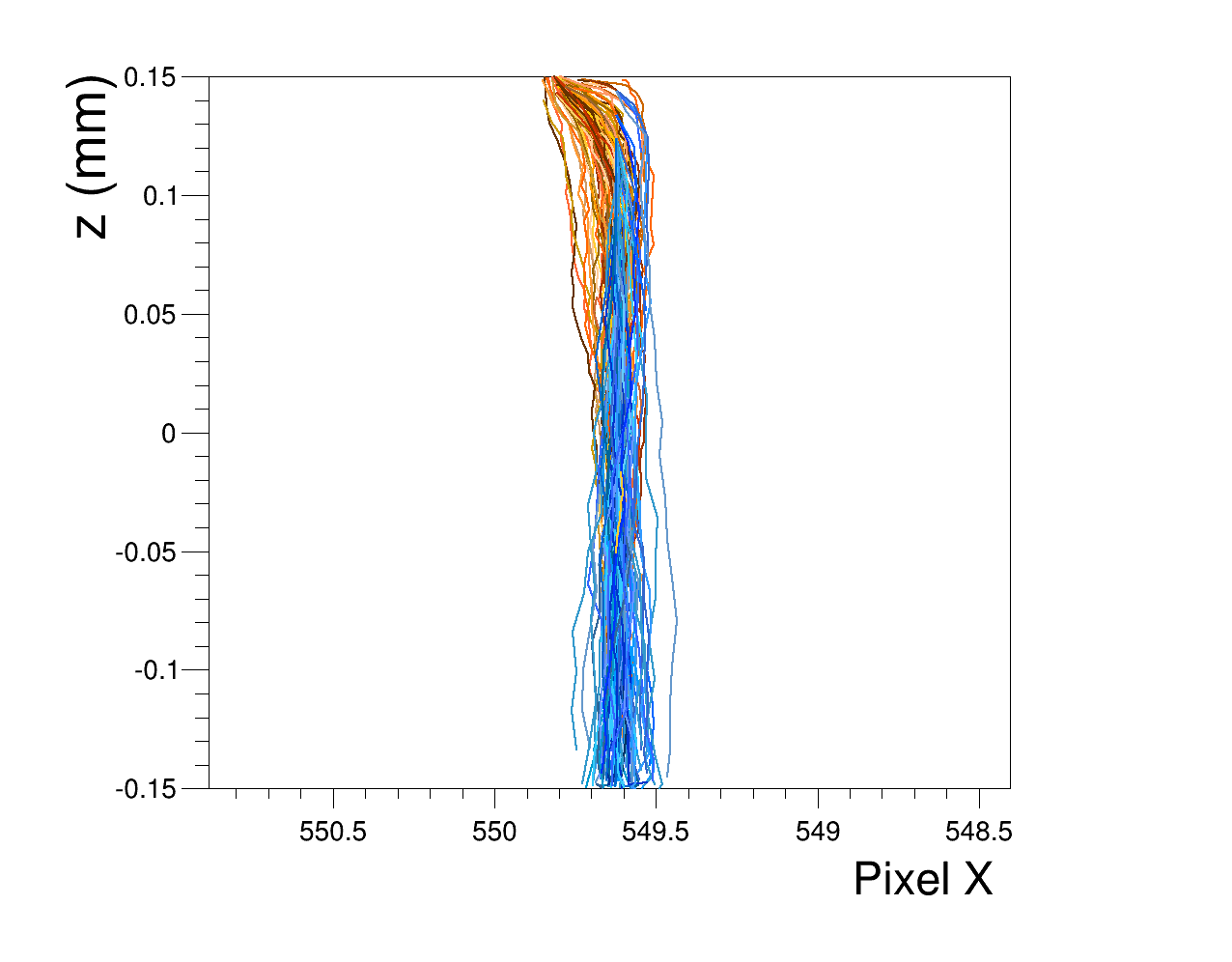}}
\subfloat{\includegraphics[scale=0.1]{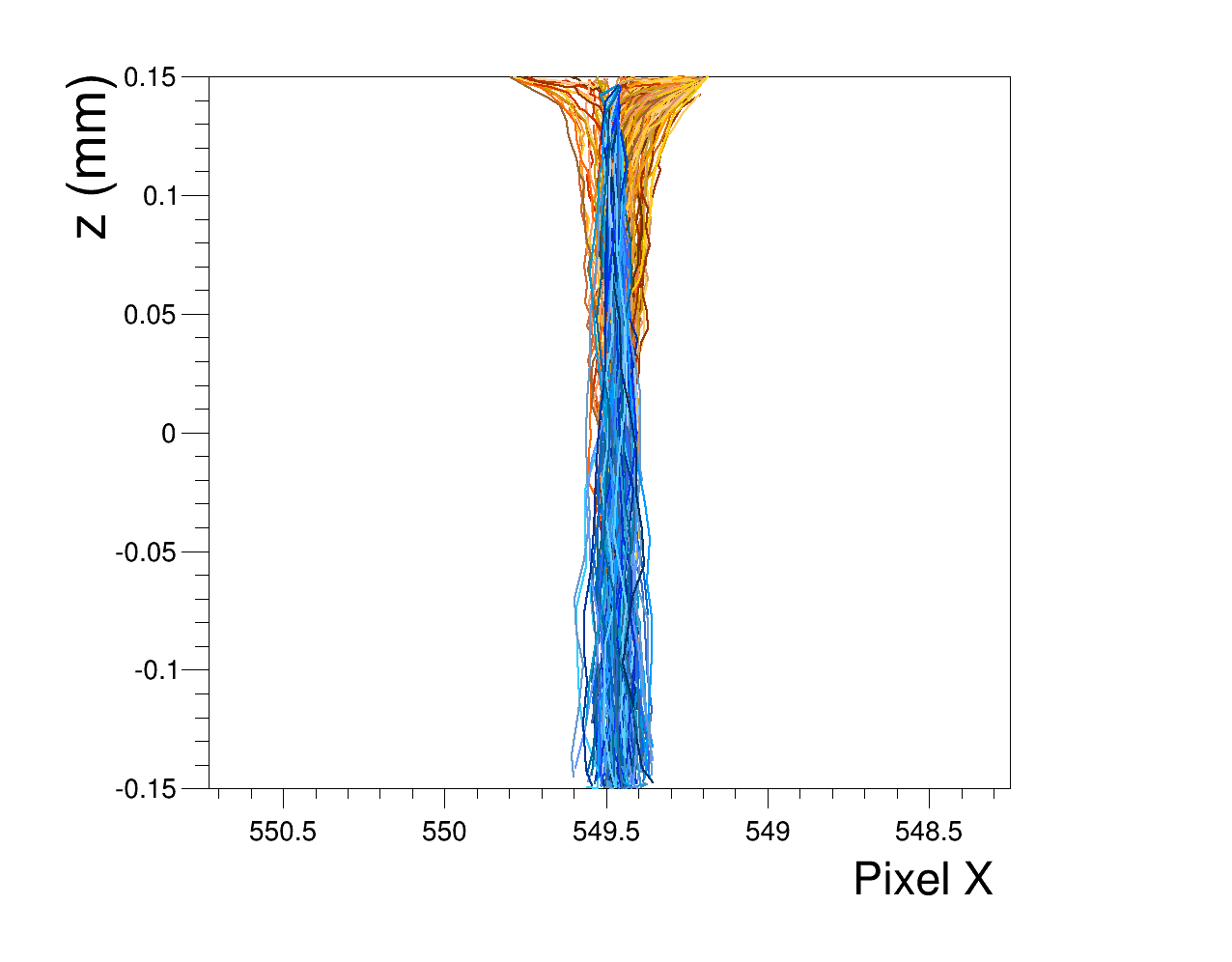}}
\subfloat{\includegraphics[scale=0.1]{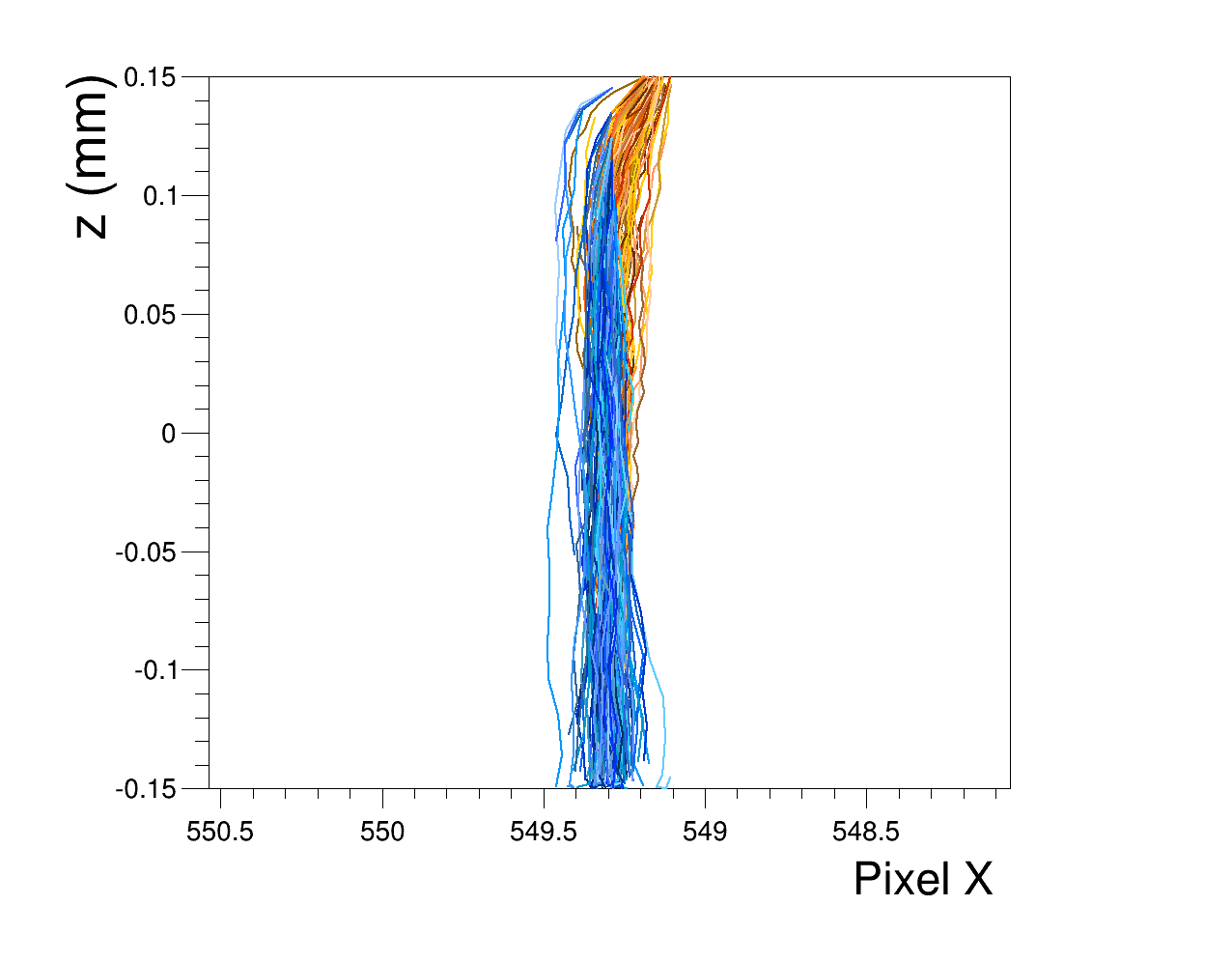}} 
\caption{When particles under perpendicular incidence strike different positions within a pitch width, the movement of charge carriers is observed. Electrons are identified by their distinctive blue coloration, while holes are represented using various shades of orange.}
\label{allpix_fig}
\end{figure}

In addition, $\rm Allpix^{2}$ offers a wide range of modules such as the charge carrier lifetime \& recombination module, the trapping and detrapping of charge carriers module, and impact ionization models. These modules can be customized by users, greatly assisting in the simulation of more accurate and precise data. $\rm Allpix^{2}$ also generates a diverse range of output plots, including the capability to create 3D GIF animations of charge carriers. Figure \ref{allpix_fig} provides a clear visualization of the movement of charge carriers within the detector. It illustrates the dispersion of charge carriers towards both sides when a particle interacts with the region between two pitches. These visual representations offer a comprehensive depiction of the internal dynamics of charge carriers in the detector, enabling a better understanding of their behavior.

\subsection{Charge sharing and signal digitization}
\begin{figure}[!ht]
\centering
\subfloat[\label{fig:DAC_0}]{\includegraphics[scale=0.22]{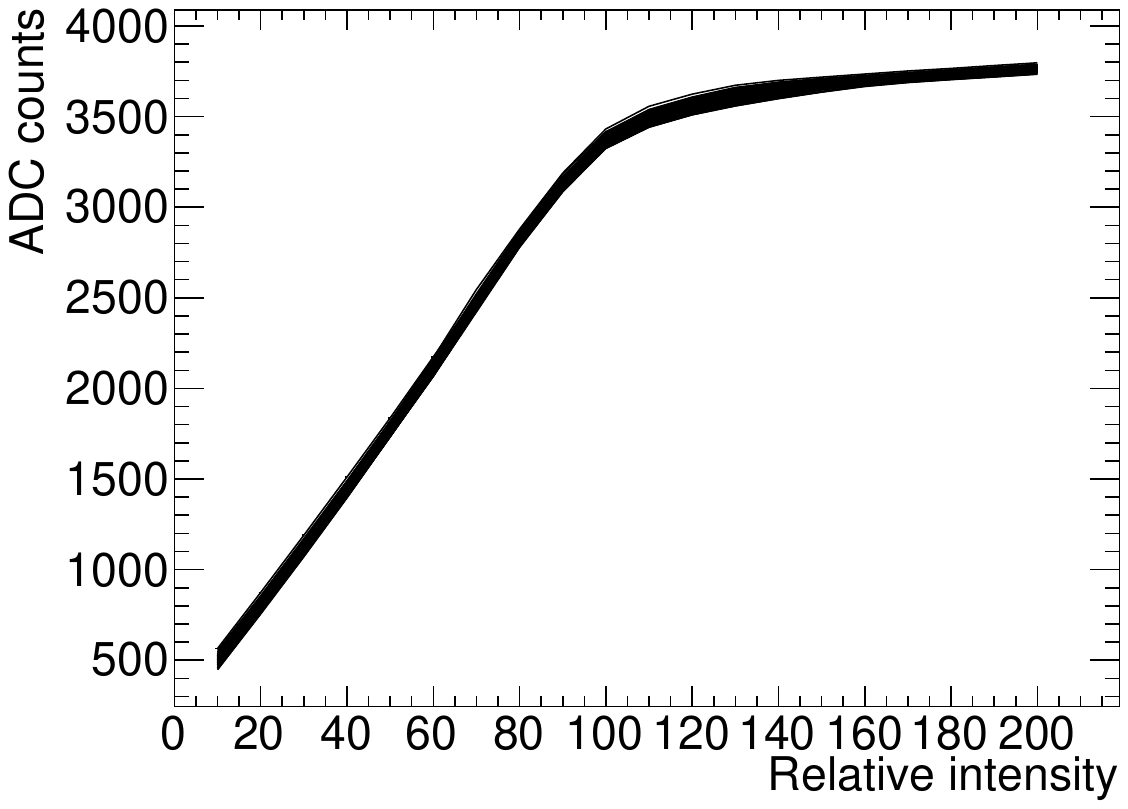}}
\subfloat[\label{fig:DAC_1}]{\includegraphics[scale=0.22]{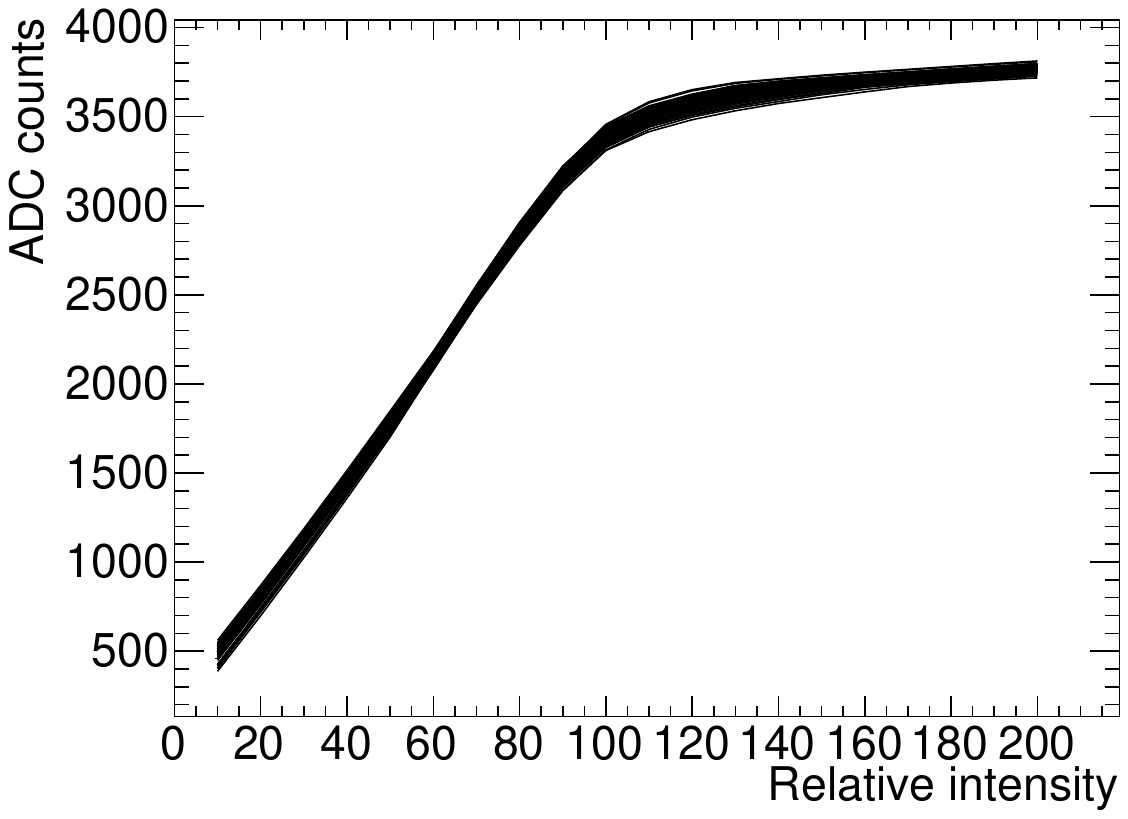}}
\subfloat[\label{fig:DAC_2}]{\includegraphics[scale=0.22]{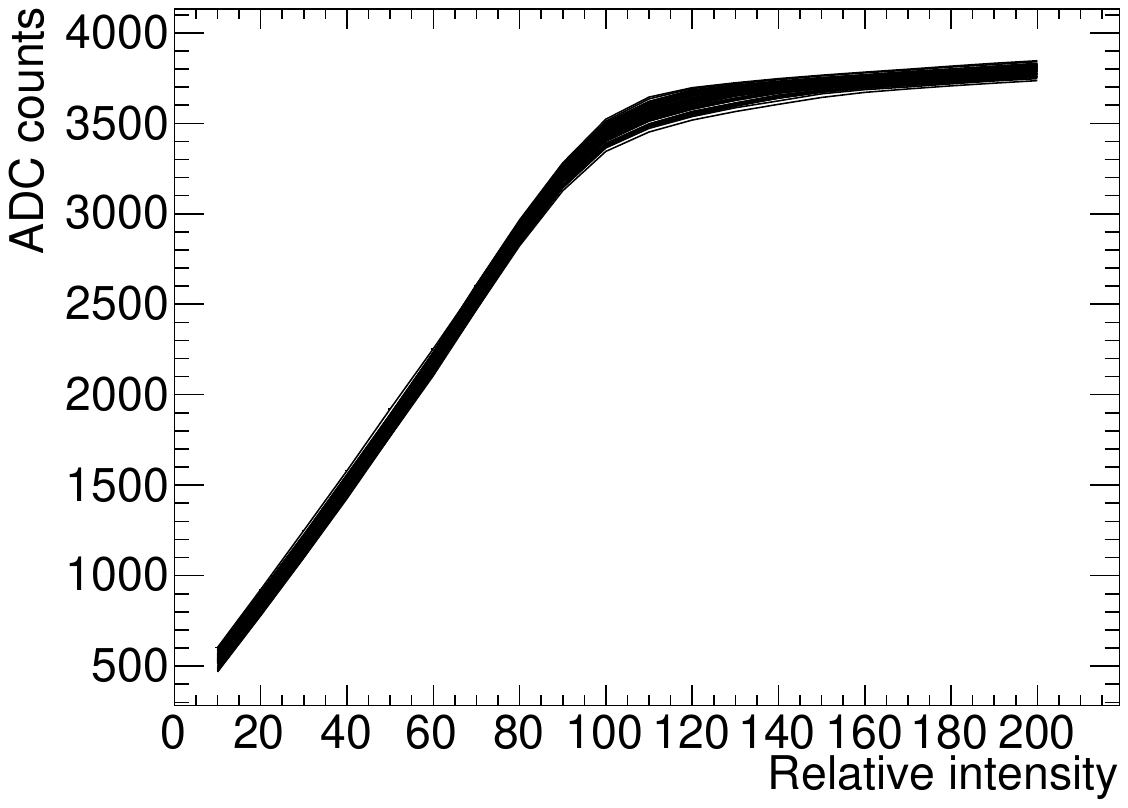}} 
\caption{The electronic linear calibration data for all channels of three randomly chosen VA140 chips are obtained. The input pulse signals with gradient amplitudes are generated by the DAC chip built into the VA140.}
\label{DAC}
\end{figure}

During the process of digitization, a simple digitization module translates the collected charges into a digitized signal proportional to the input charge. In ground testing, the ASICs demonstrated good linearity, with an average integral nonlinearity (INL) of less than 1.5\% \cite{va_test}. Figure \ref{DAC}  shows that on-orbit data displays high linearity as well in the unsaturated zone; gain calibration for each VA and alignment were finished in the earlier work\cite{stk_gain,my_align}. Due to the excellent consistency of on-orbit data, we have established the same qdc\_slope for all channels. The qdc\_slope is defined as the slope of the charge-to-digital converter (QDC) calibration, measured in electrons per ADC unit. The ADC has a 12-bit resolution. Simulation results were ultimately stored in the ROOT format for further analysis.

\begin{figure}[!h]
\centering
\includegraphics[scale=0.32]{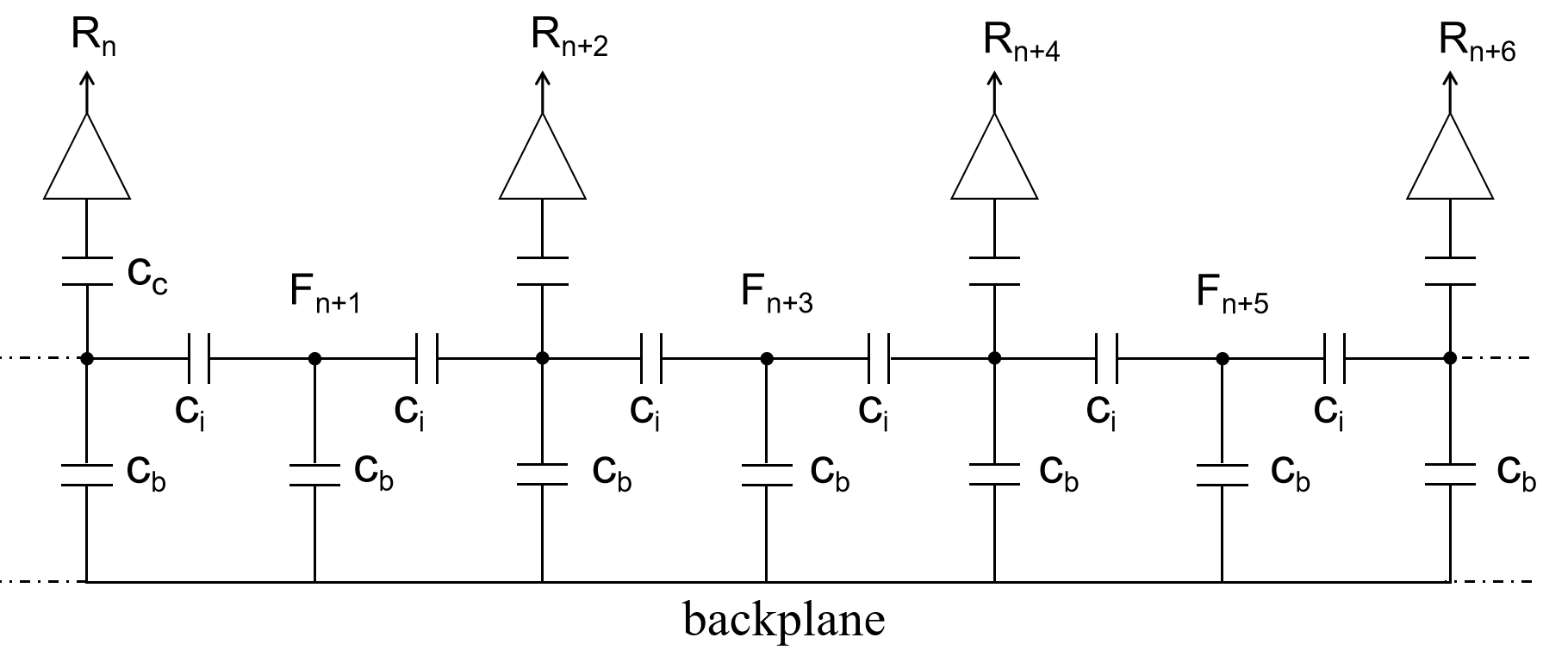}
\caption{The equivalent circuit model of the sensor}
\label{fig_spice}
\end{figure}

The sensor of the STK is commonly represented as a network of capacitors, as described in Ref.~\cite{charge_qiao}. In Figure \ref{fig_spice}, the strips are categorized into two types: readout strips, denoted by R and connected to the VA, and floating strips, denoted by F and not connected to the VA. The channel numbers are indicated by subscripts.The capacitors $C_{b}$, $C_{c}$, and $C_{i}$ represent the strip-to-backplane capacitance, the coupling capacitance, and the inter-strip capacitance between the implantation strips, respectively.

Typically, TCAD simulation software can be used to simulate the coupling capacitance between different pixels (strips). The obtained capacitance values can be incorporated into the SPICE model~\cite{spice} to obtain the charge distribution coupling matrix between different strips. $\rm Allpix^{2}$ provides extensive interfaces to incorporate this information into simulated data, including the ability to account for non-uniform coupling capacitance.

As $\rm Allpix^{2}$ only provides the generated charge signal of each strip, we need incorporate a noise component as determined with on-orbit data into the corresponding simulated readout strips, and then independently handle the remaining digitization process based on the specific characteristics of our detector.

To begin with, we need to import a sharing coefficient that transfers the signal from the floating strip to the adjacent readout strip. One way to obtain the sharing coefficient is through the SPICE model. By importing the relevant capacitance values into the model and injecting charges at various points, we can determine the charge coefficients for different strips. Even if an actual detector is not available, the capacitance can be numerically calculated as described in \cite{cap}.

Another method involves selecting vertical entries from beam experiments or on-orbit data and quantifying the signal ratio obtained from different strips. In this case, the sharing process utilizes results obtained from on-orbit data. By incorporating the obtained sharing coefficients, we can accurately simulate the hit-sharing process.

To ensure reliable performance, it is important to have consistent channel indices and corresponding noise in the simulation compared to the on-orbit data. Due to data transmission limitations, the satellite only downloads baseline data once per day. Since the DAMPE satellite operates stably \cite{stk_gain}, we randomly selected baseline data from a single day and calculated the noise for all channels using the method described in reference \cite{noise_alg}. During the digitization process, this calculated noise is added to the corresponding channels in the simulated data. In addition, for a few channels with poor performance, the corresponding channels are masked during the simulation process to account for their absence.

\begin{figure}[!h]
\centering
\includegraphics[scale=0.6]{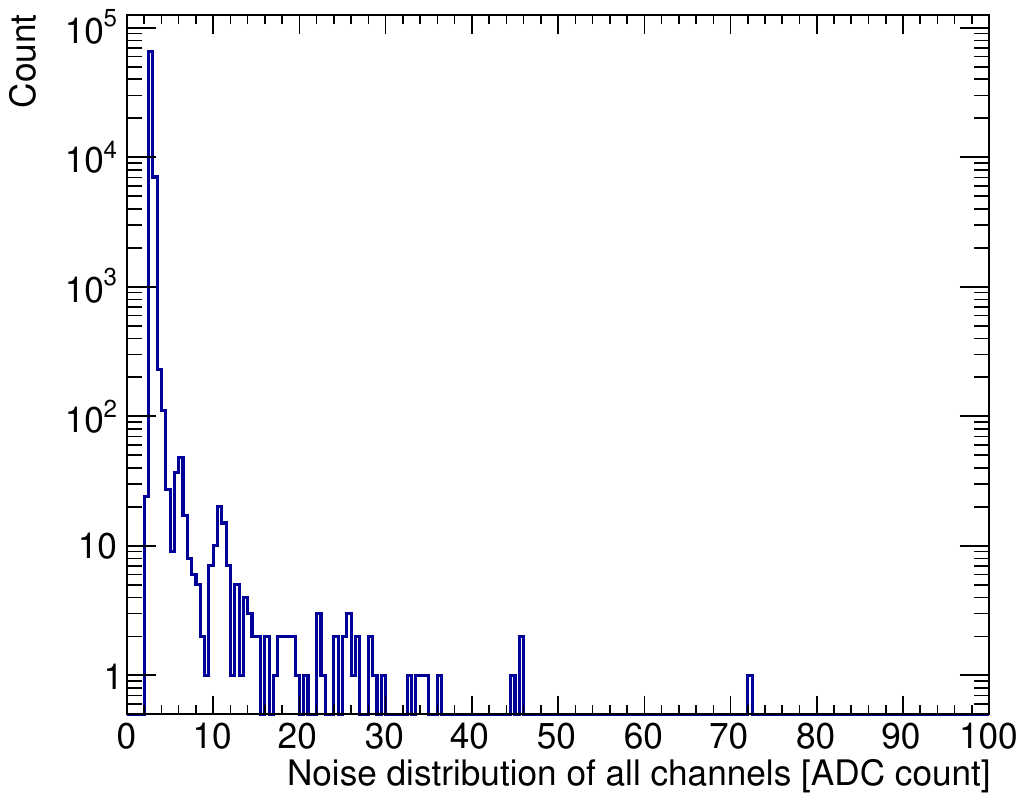}
\caption{Distribution of imported noise in simulated data.}
\label{fig_noise}
\end{figure}

The noise results from the on-orbit data are shown in Figure \ref{fig_noise}, where it can be observed that the noise in the majority of channels is within 3 ADC. The proton MIP collected by the readout strip has an approximate value of 55 ADC.

\subsection{Cluster construction and track reconstruction}
In principle, when a charged particle passes through the silicon microstrip detector, it produces a series of consecutive hit points, referred to as clusters. Cluster formation involves searching for seed channels with a signal-to-noise ratio (S/N) greater than 4 and including neighboring channels with S/N greater than 1.5. The cluster construction is essential for determining the incident position and particle charge. A cluster's information includes its size, seed channel, total ADC count, CoG (Center-of-Gravity), $\eta$ parameter, and other relevant quantities that describe tracking performance. The CoG is calculated using the following equation:

\begin{equation}\label{chi2}
CoG = \frac{\sum_{i}^{n}\frac{S_{i}}{N_{i}}\cdot pos_{i} }{\sum_{i}^{n}\frac{S_{i}}{N_{i}}},
\end{equation}

where $i$ represents the index number of a channel, $S_{i}$ and $N_{i}$ are the signal and noise values, respectively, and $pos_{i}$ is the coordinate of the $i$-th channel. The CoG of the cluster is then used to create the track seed. The $\eta$ parameter is defined as the CoG of only two channels: the channel with the highest amplitude and its neighboring channel with a larger coordinate. $\eta$ is useful for determining the relative incident position of particles.

\begin{figure}[!h]
\centering
\subfloat[\label{fig:track}]{\includegraphics[scale=0.34]{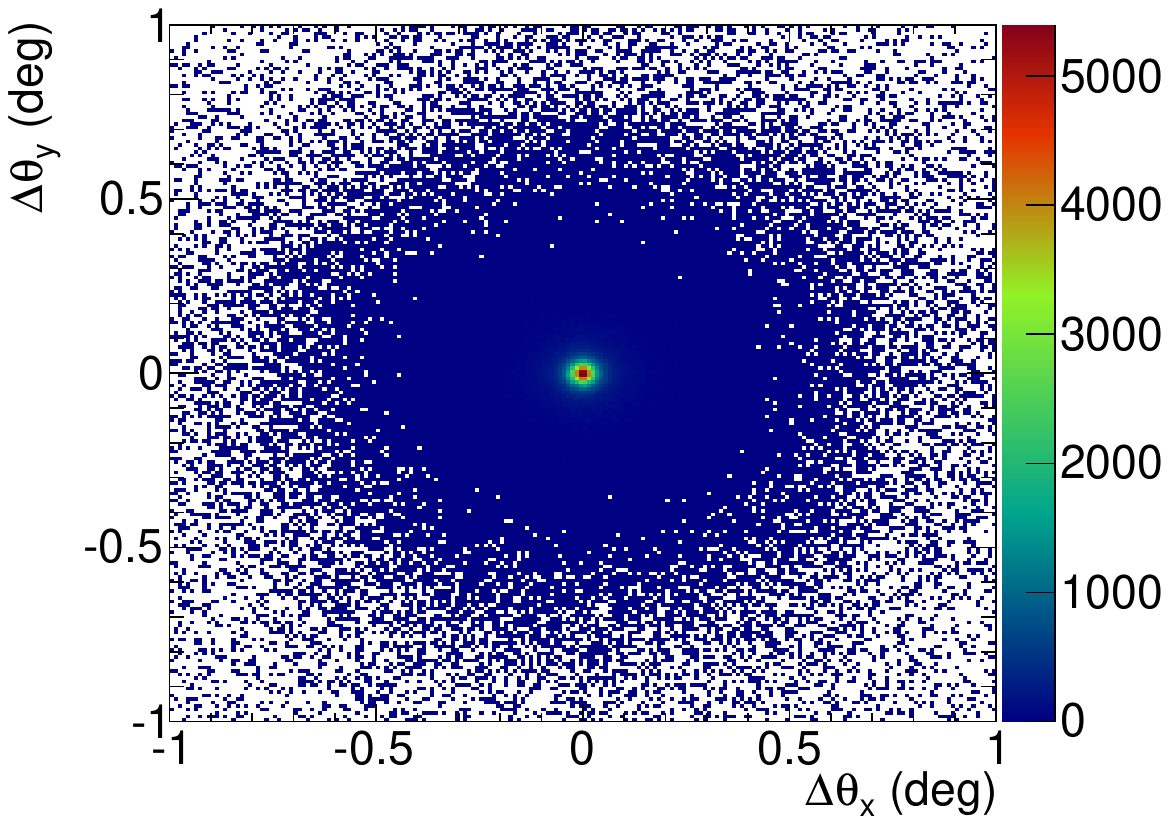}}
\subfloat[\label{fig:res}]{\includegraphics[scale=0.34]{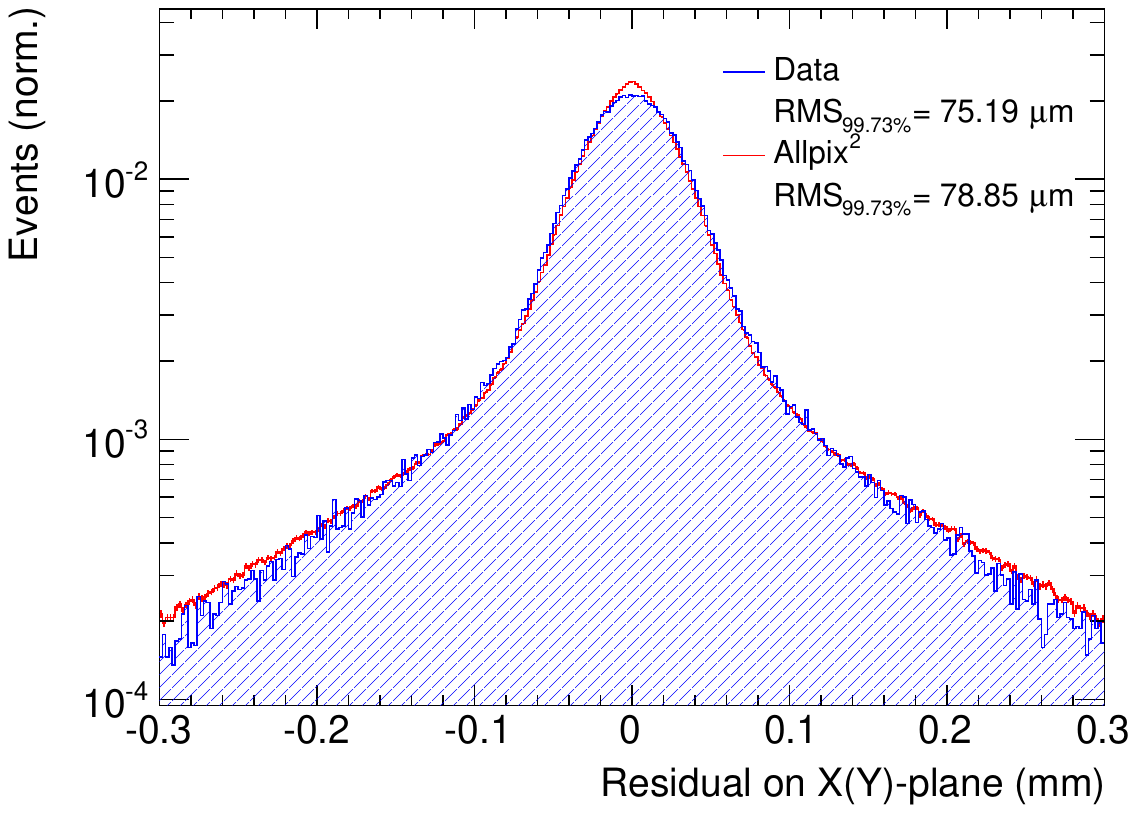}}
\caption{The left figure illustrates the deviations between the reconstructed tracks and the primary ones, while the right figure shows the residuals in the X and Y directions for both the data and simulation.}
\label{fig:track_com}
\end{figure}

The track is reconstructed using a custom implementation of the Kalman filter algorithm \cite{kalman}, similar to the approach used with on-orbit data. Figure \ref{fig:track} illustrates the spatial resolution of the simulated detector for proton MIPs using the reconstructed cluster position and the Monte Carlo truth information from the primary particle. Figure \ref{fig:res} shows the residual in the X and Y directions, which represents the difference between the incident particle's impact point obtained from the on-orbit data and the cluster position reconstructed in the simulation. The width of the residual is calculated as the root mean square (RMS) of the distribution, evaluated for the middle 99.73\% of the histogram.

\section{Simulation validation}\label{sed:comparison}
To validate the performance of the developed framework, the simulation results are compared with the beam test data as well as the on-orbit data of DAMPE experiment.

\subsection{ Validation with test-beam data}
\begin{figure}[!h]
\centering
\subfloat[\label{fig:beam_cs_all}]{\includegraphics{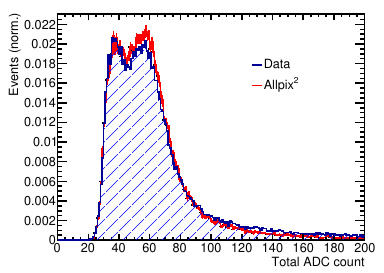}}
\subfloat[\label{fig:beam_cs_size}]{\includegraphics{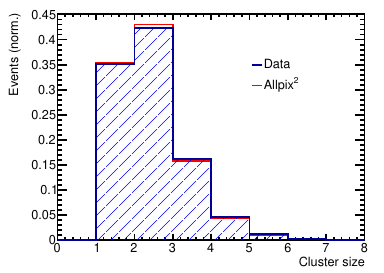}}
\caption{Fig.\ref{fig:beam_cs_all} shows the cluster ADC distribution for simulation and beam-test data, and Fig.\ref{fig:beam_cs_size} represents the cluster size.}
\label{beam-test}
\end{figure}

The simulation is first compared to data obtained from a beam test conducted at the CERN-SPS accelerator using a 400 GeV proton beam in June 2015. The beam test data were reprocessed using the latest reconstruction and selection algorithms, which are identical to those used in the analysis of on-orbit data \cite{dampe:beam}. Since the beam direction is almost perpendicular to the detector, we selected events with reconstructed $\theta x(y)$ values less than $0.57^{\circ}$ for comparison with the simulation data, as depicted in Figure \ref{beam-test}.

\subsection{Validation with on-orbit data}\label{val_fly}
\begin{figure}[!htbp]
\centering
\includegraphics[scale=0.34]{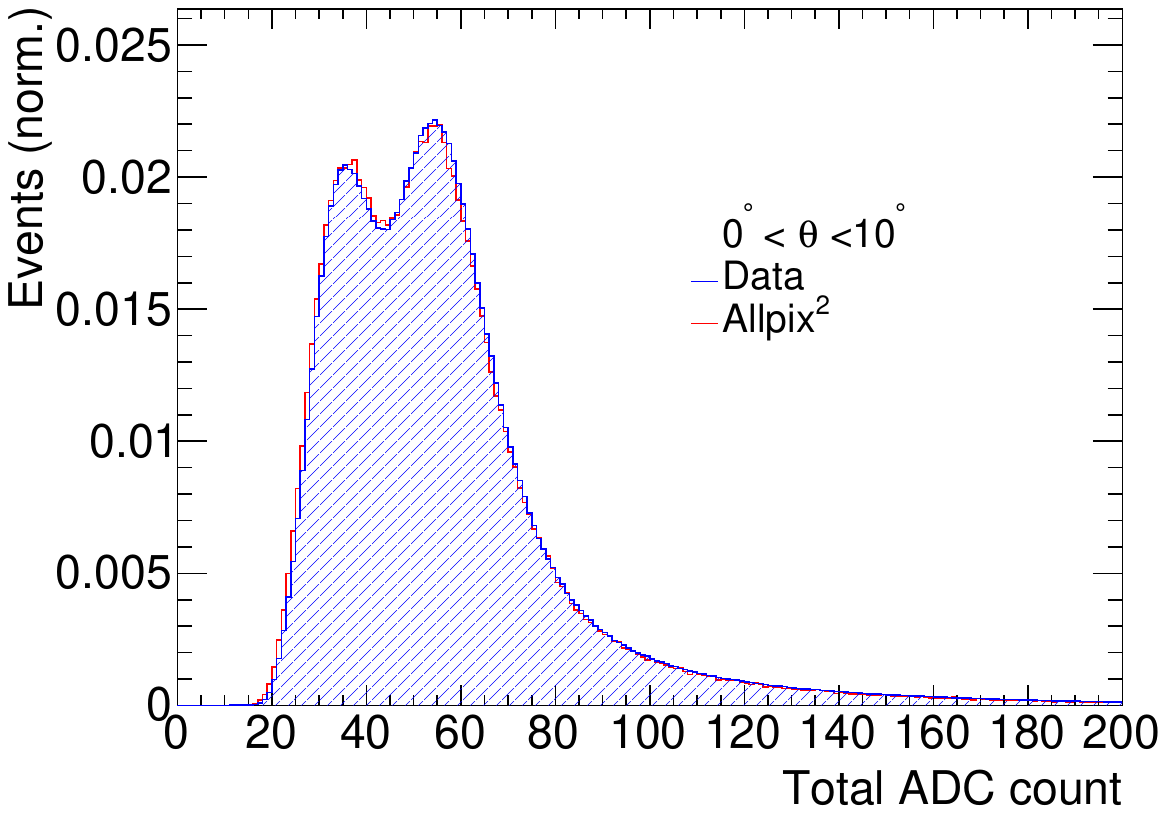}
\includegraphics[scale=0.34]{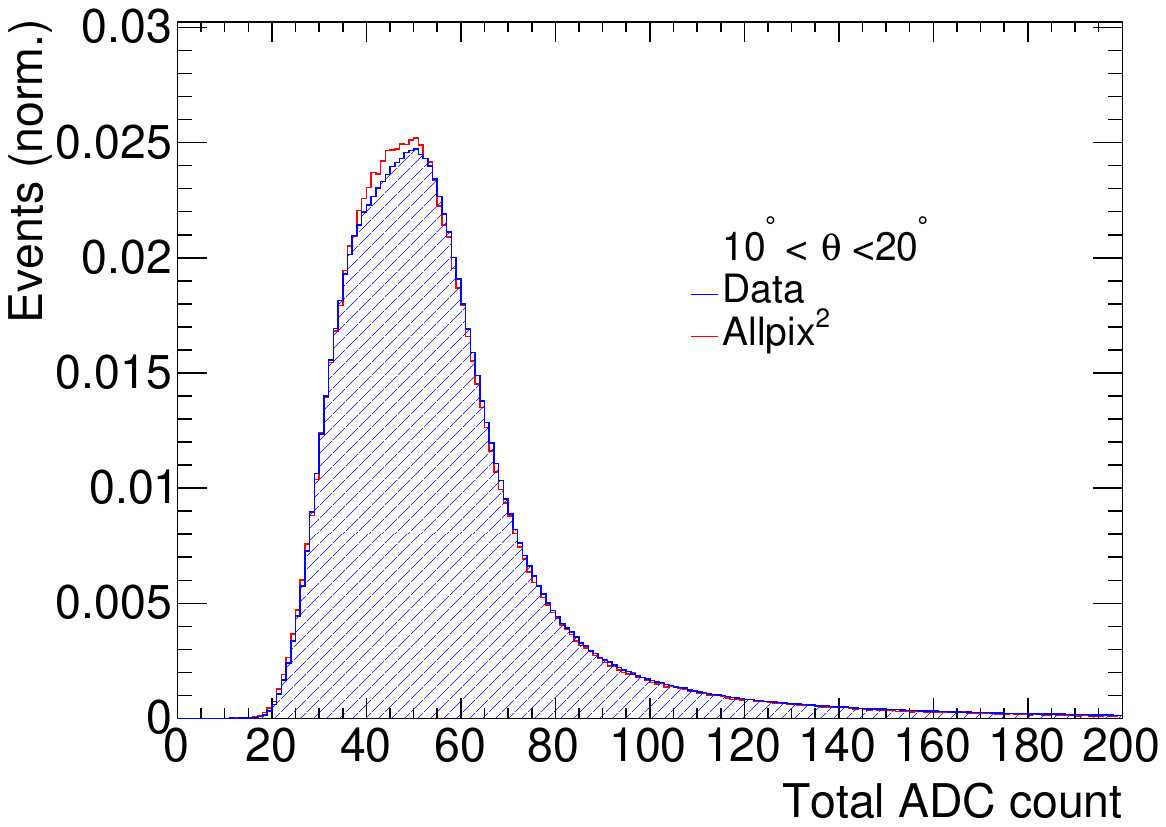}
\includegraphics[scale=0.34]{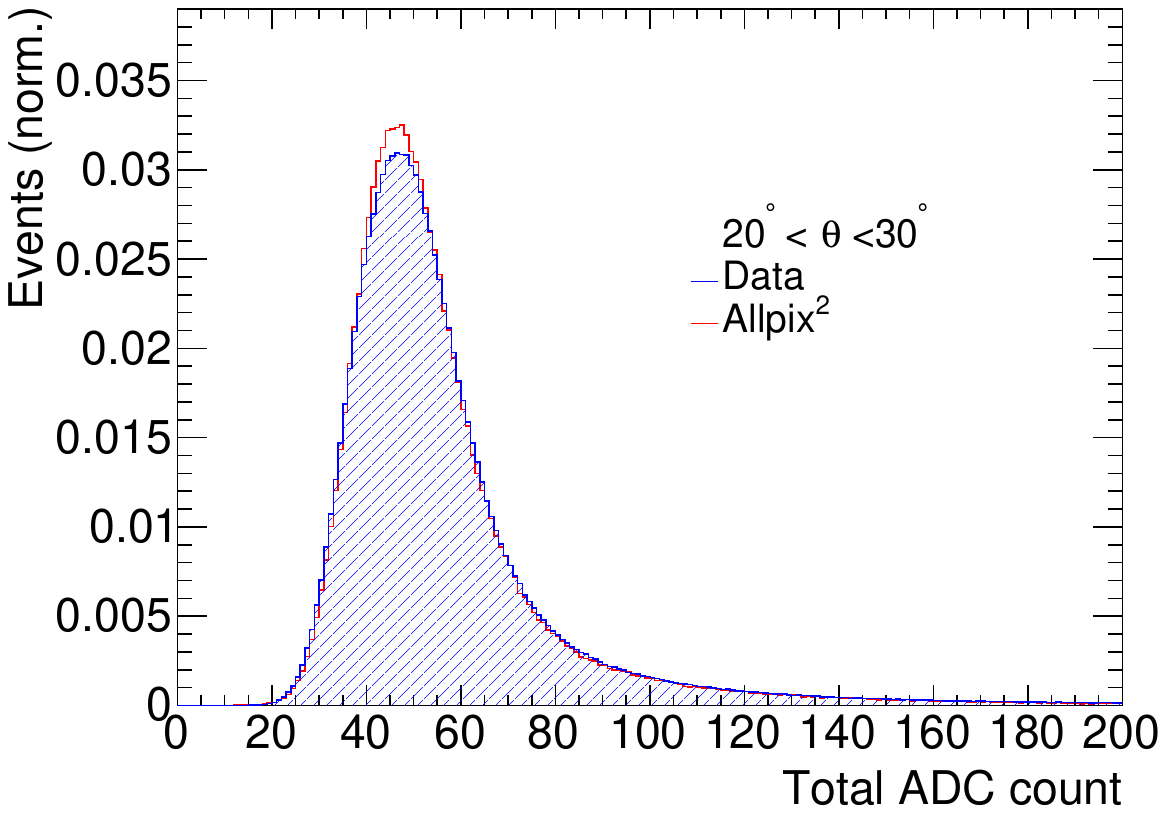}
\includegraphics[scale=0.34]{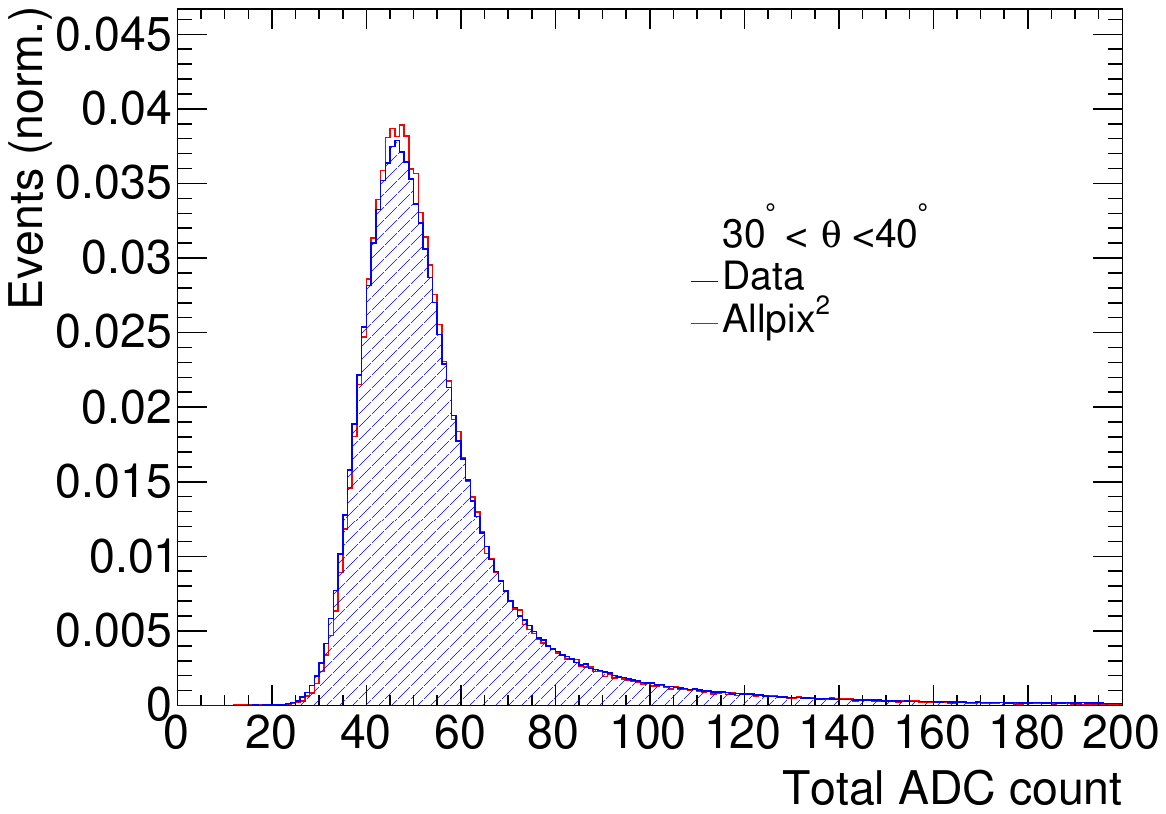}
\caption{The total ADC count comparisons of all clusters for protons in different $\theta$ bins, the normalization method used in Figures 12-15 is area normalization.}
\label{h_all}
\end{figure}

\begin{figure}[!htbp]
\centering
\includegraphics[scale=0.34]{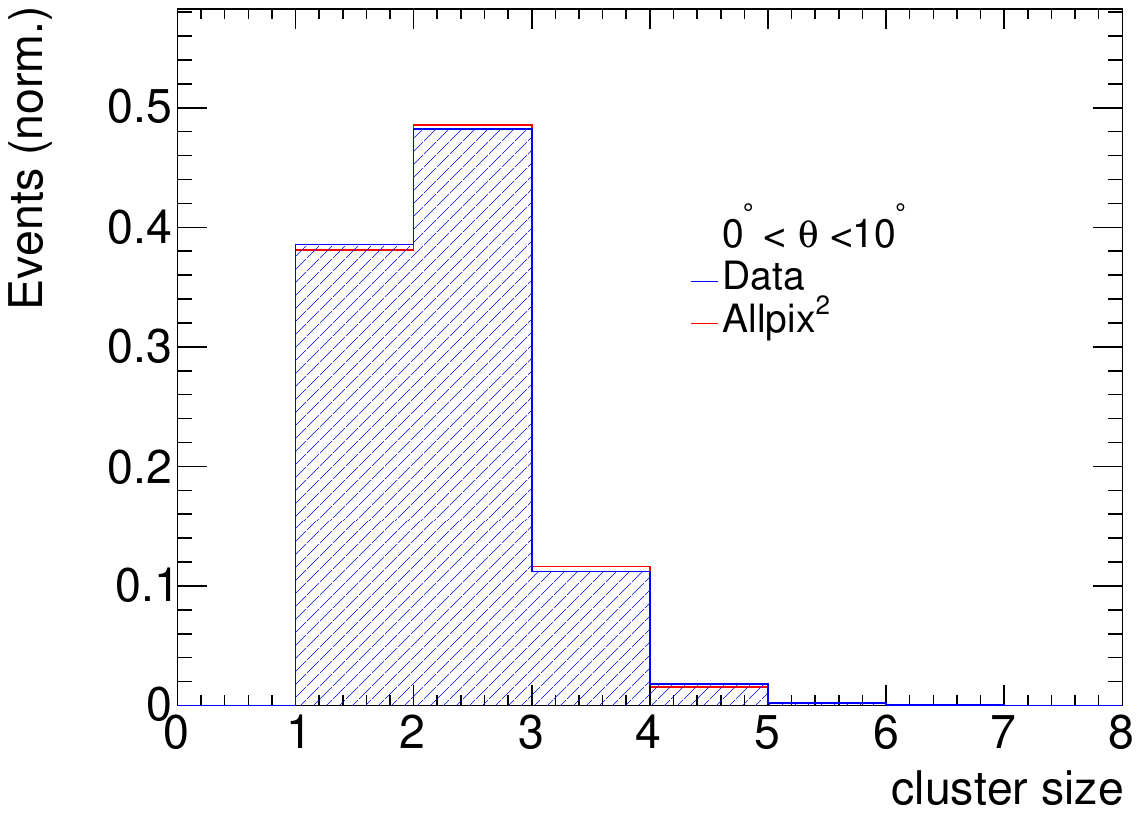}
\includegraphics[scale=0.34]{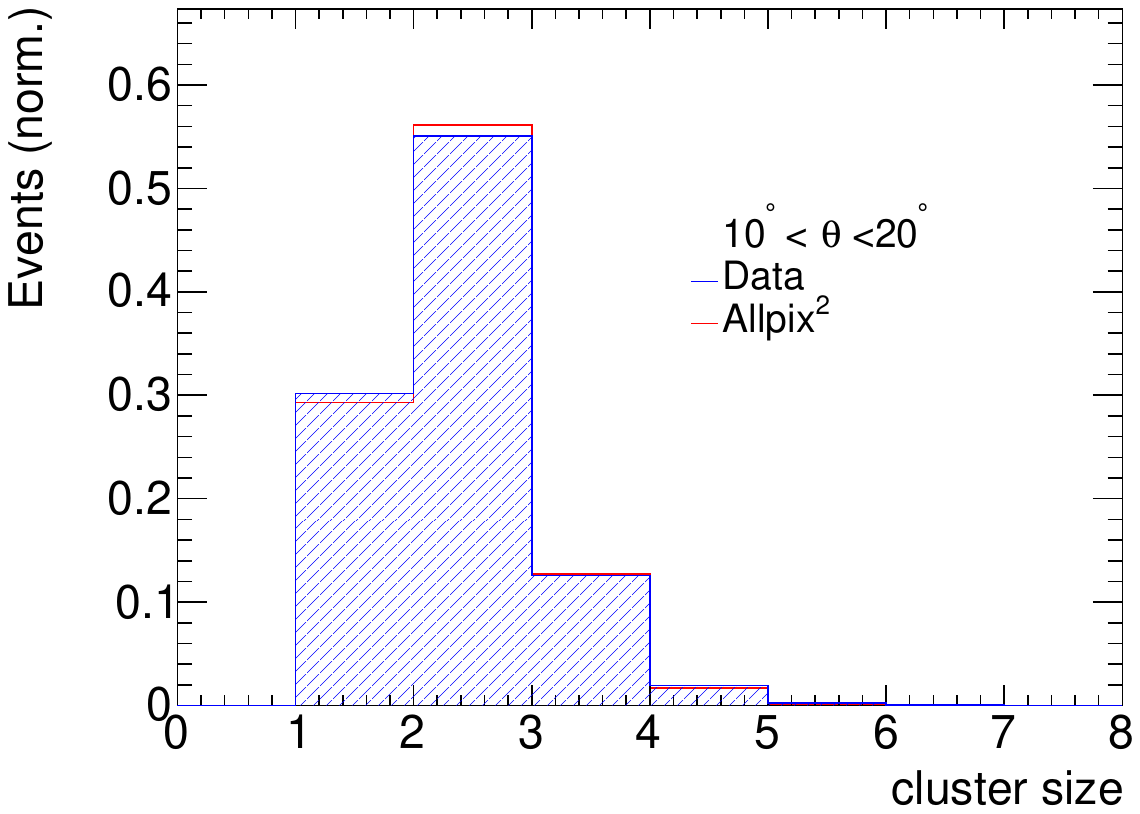}
\includegraphics[scale=0.34]{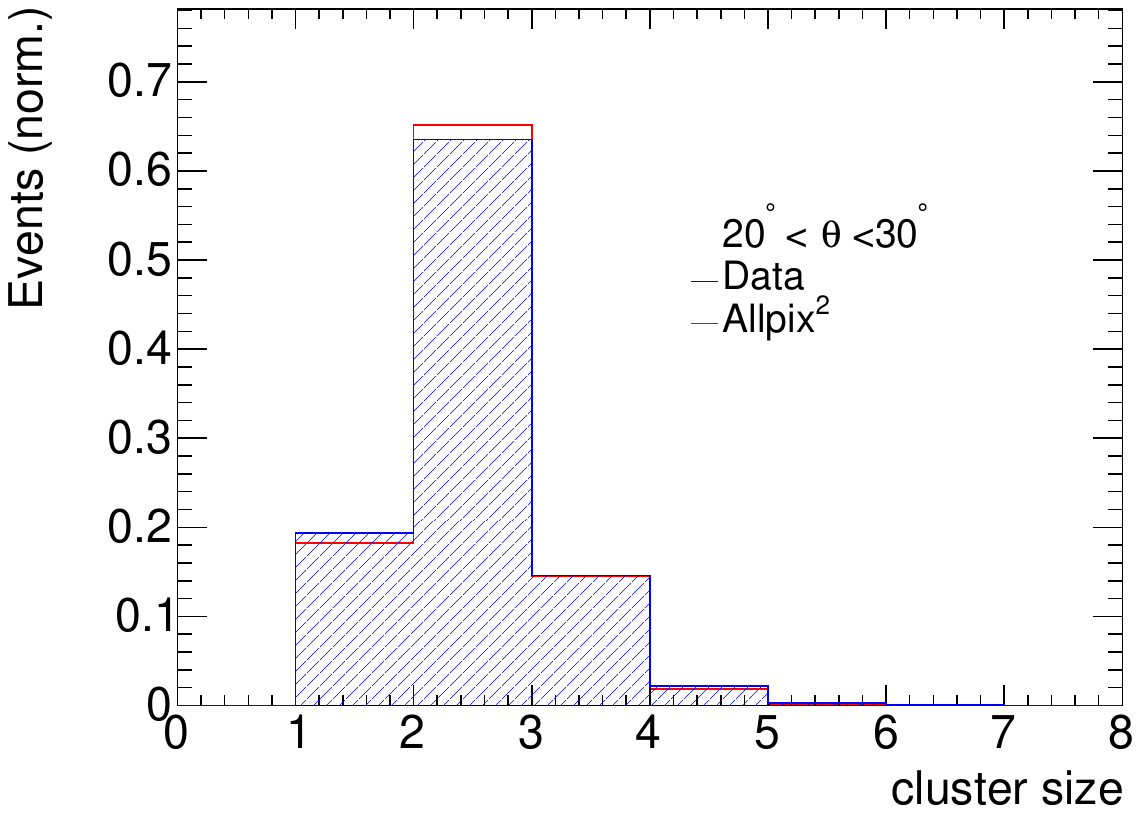}
\includegraphics[scale=0.34]{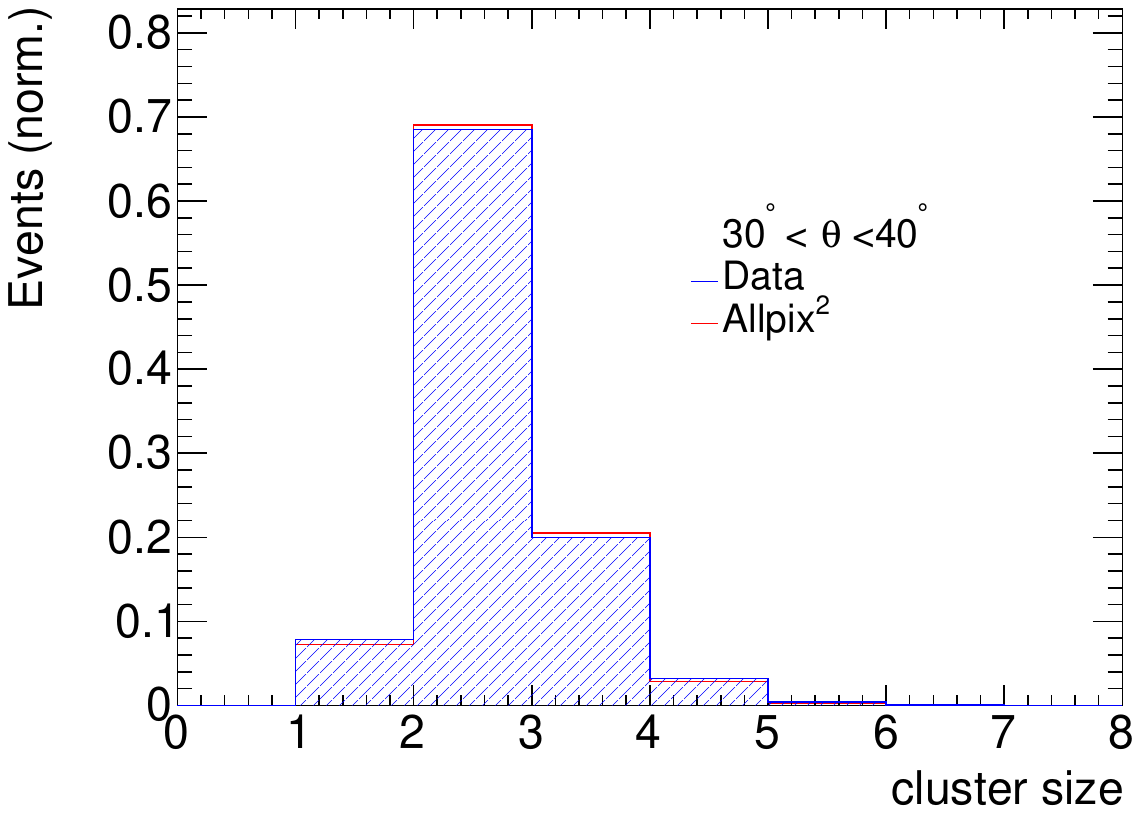}
\caption{The cluster size comparisons for protons in different $\theta$ bins.}
\label{h_num}
\end{figure}

\begin{figure}[!htbp]
\centering
\includegraphics[scale=0.34]{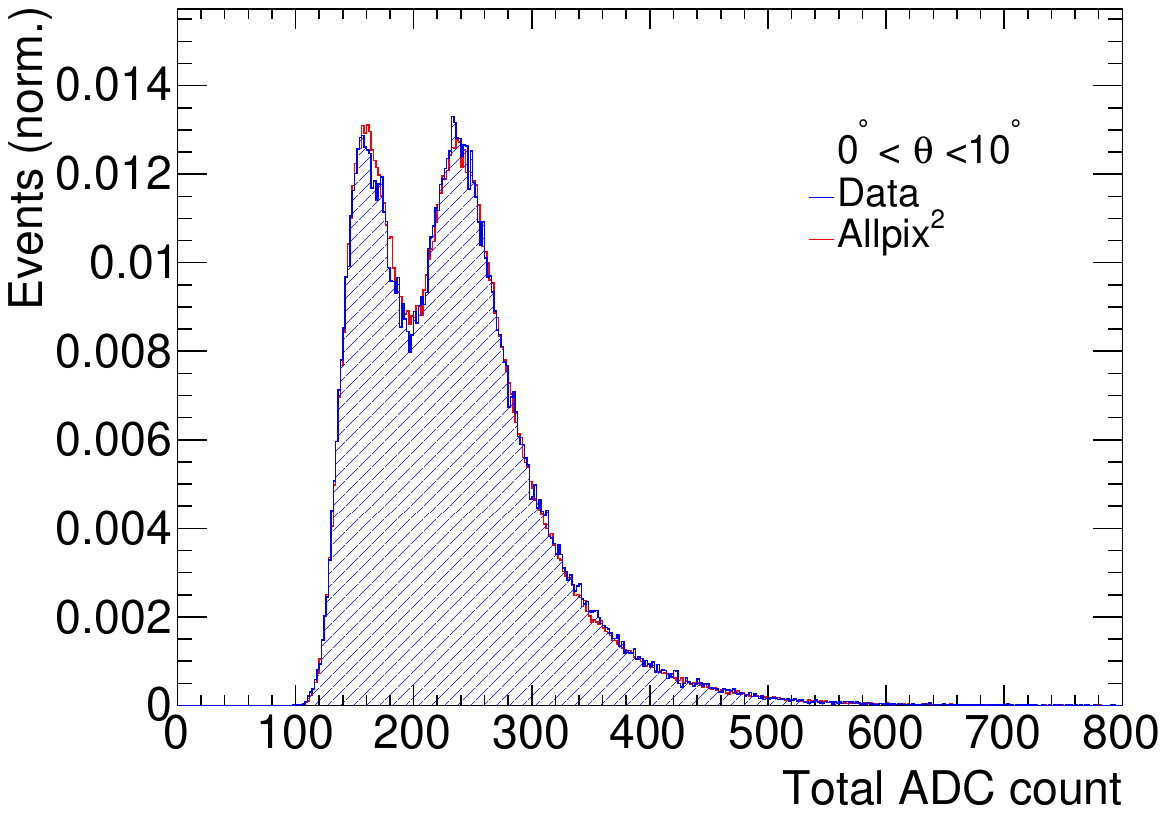}
\includegraphics[scale=0.34]{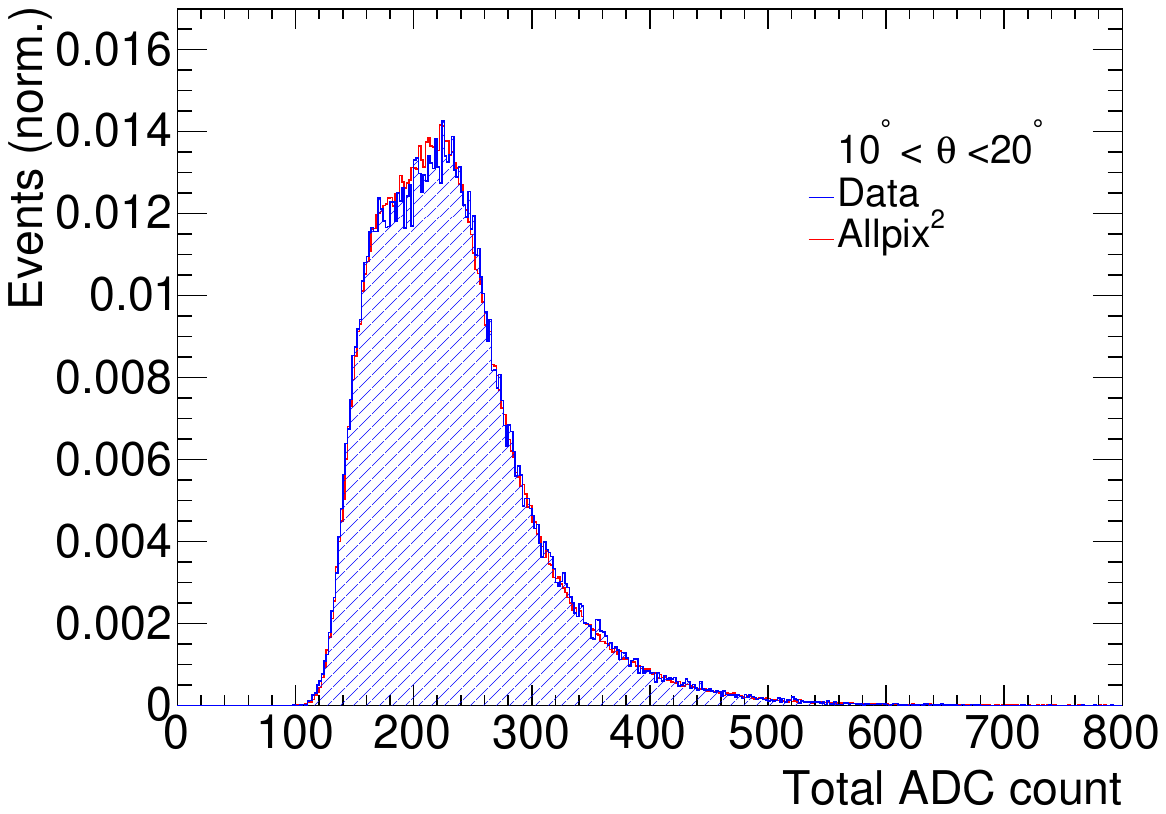}
\includegraphics[scale=0.34]{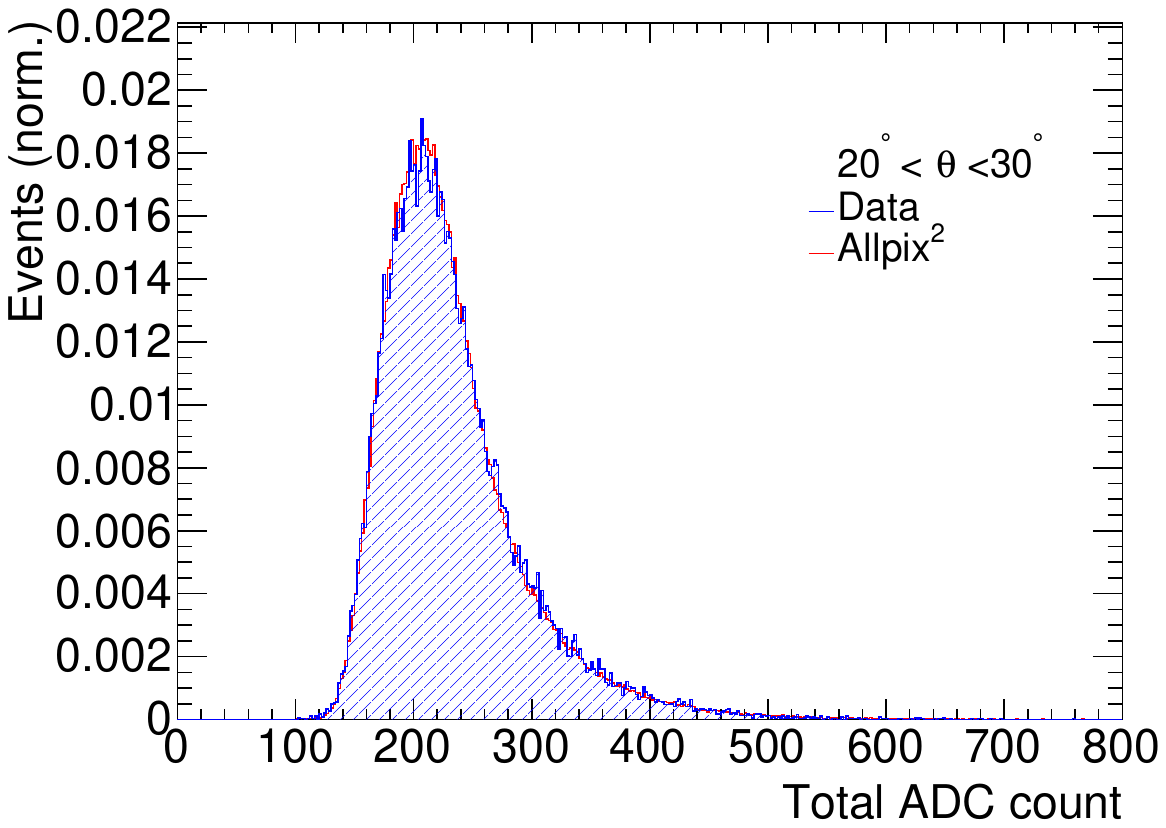}
\includegraphics[scale=0.34]{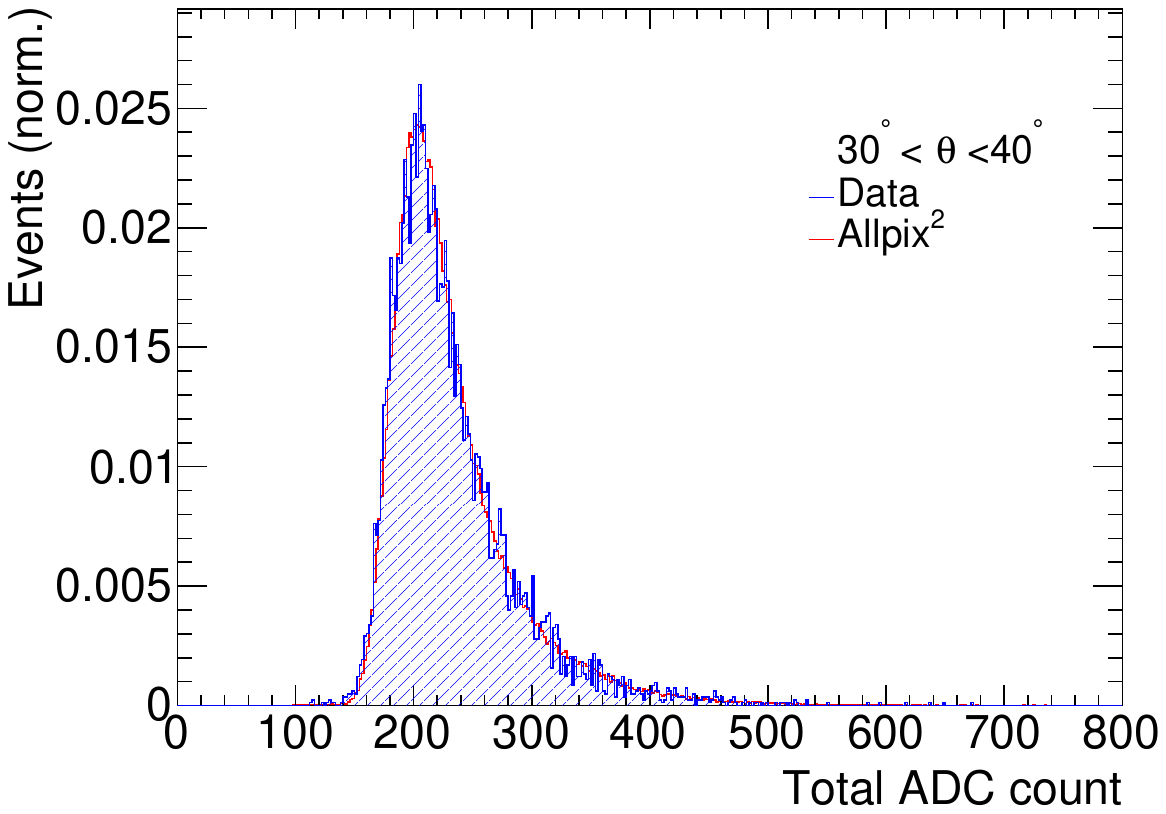}
\caption{The total ADC count comparisons of all clusters for helium nuclei in different $\theta$ bins.}
\label{he_all}
\end{figure}

\begin{figure}[!htbp]
\centering
\includegraphics[scale=0.34]{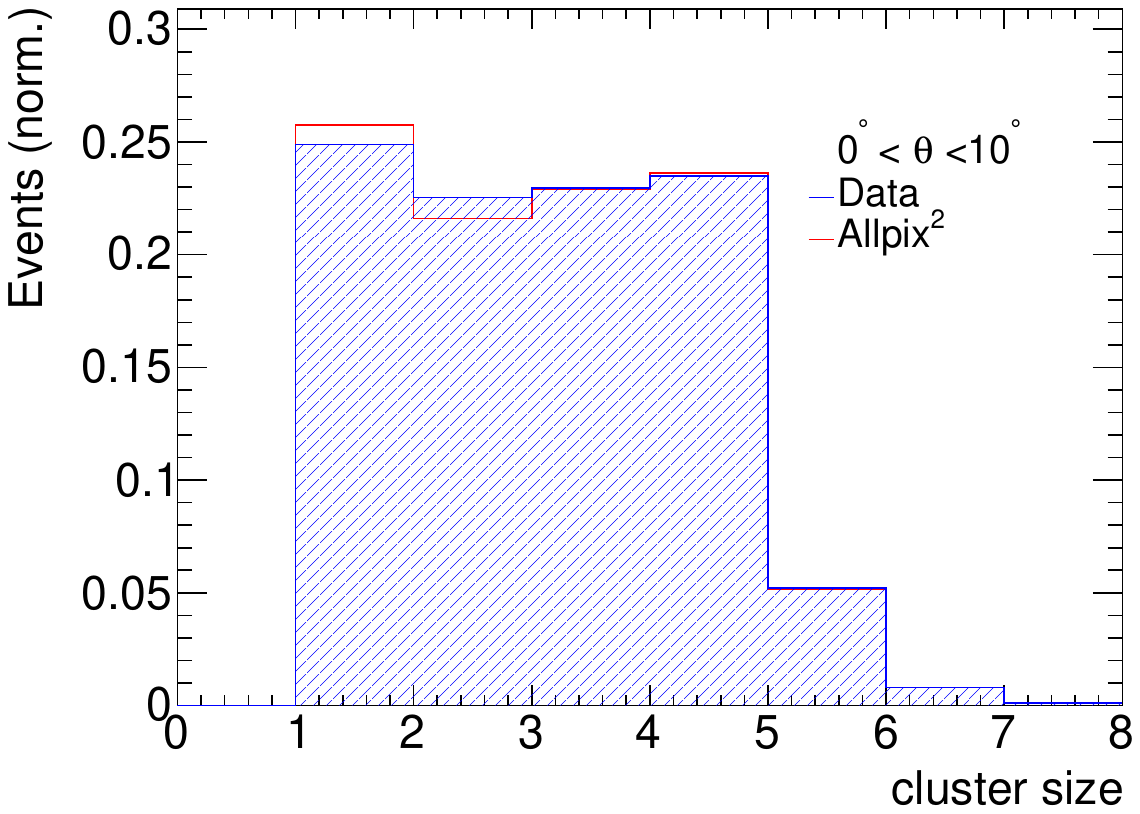}
\includegraphics[scale=0.34]{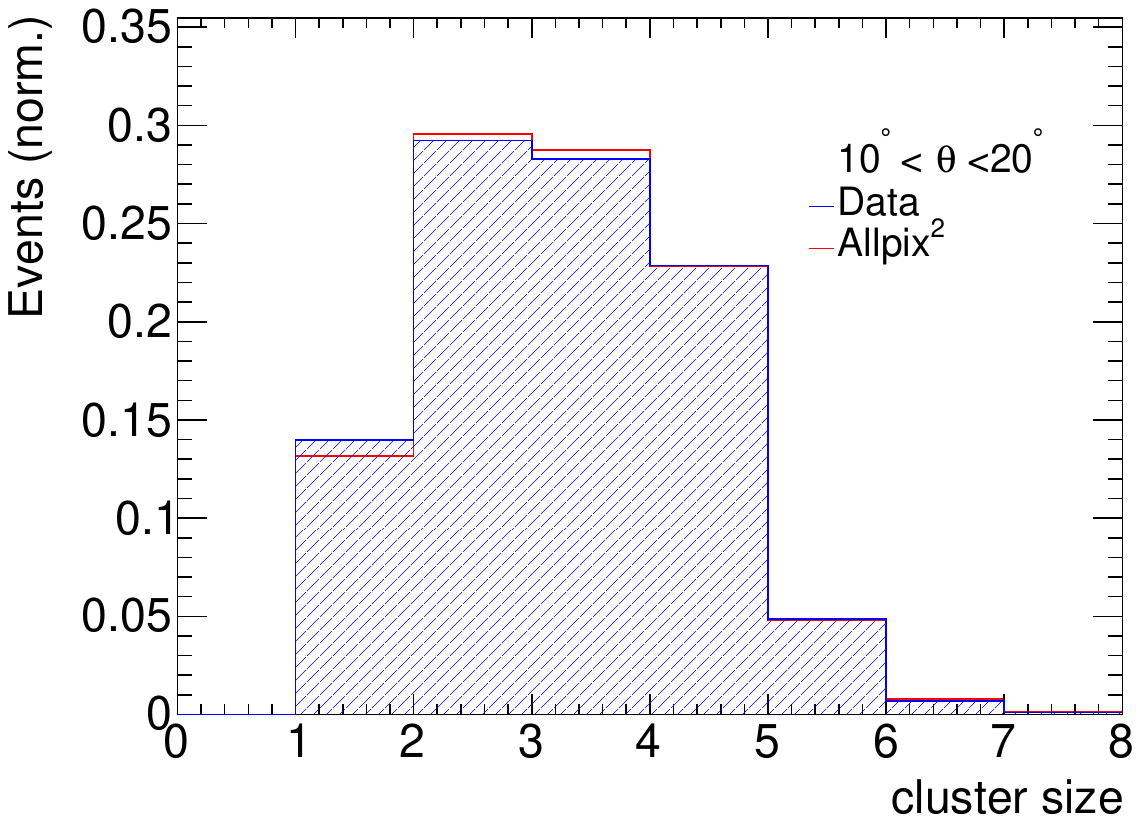}
\includegraphics[scale=0.34]{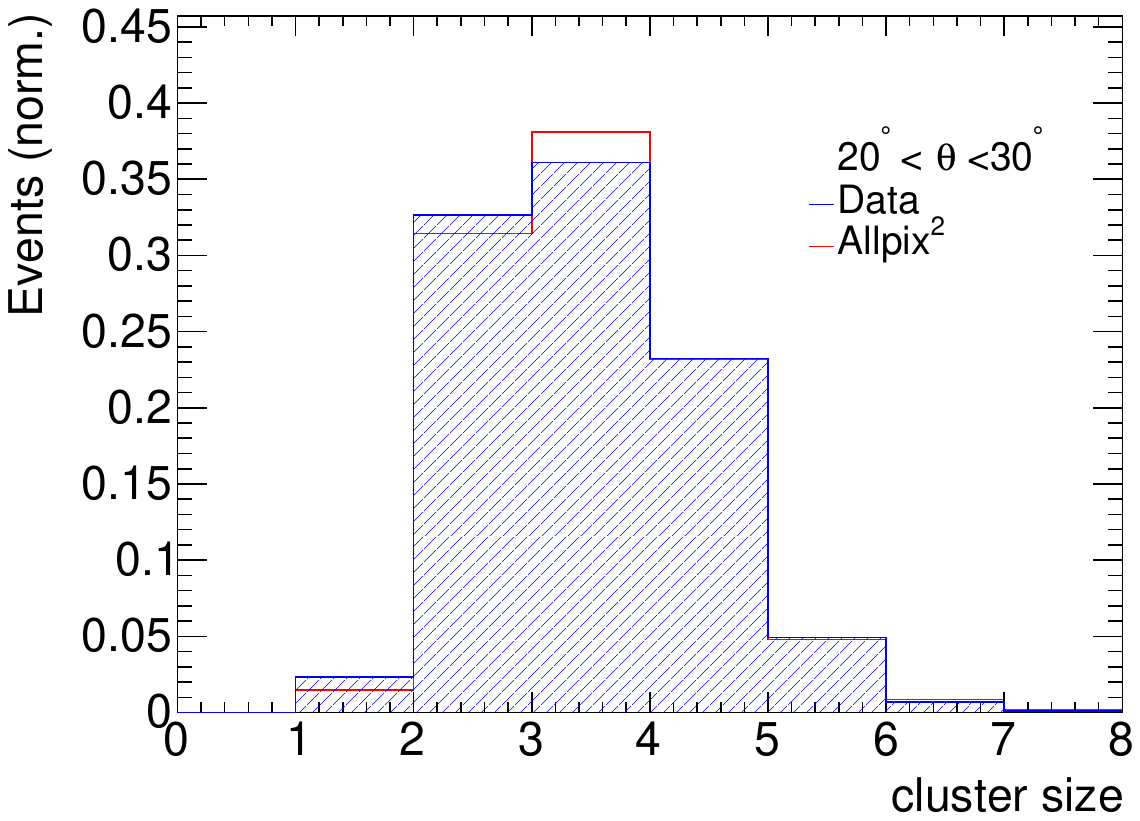}
\includegraphics[scale=0.34]{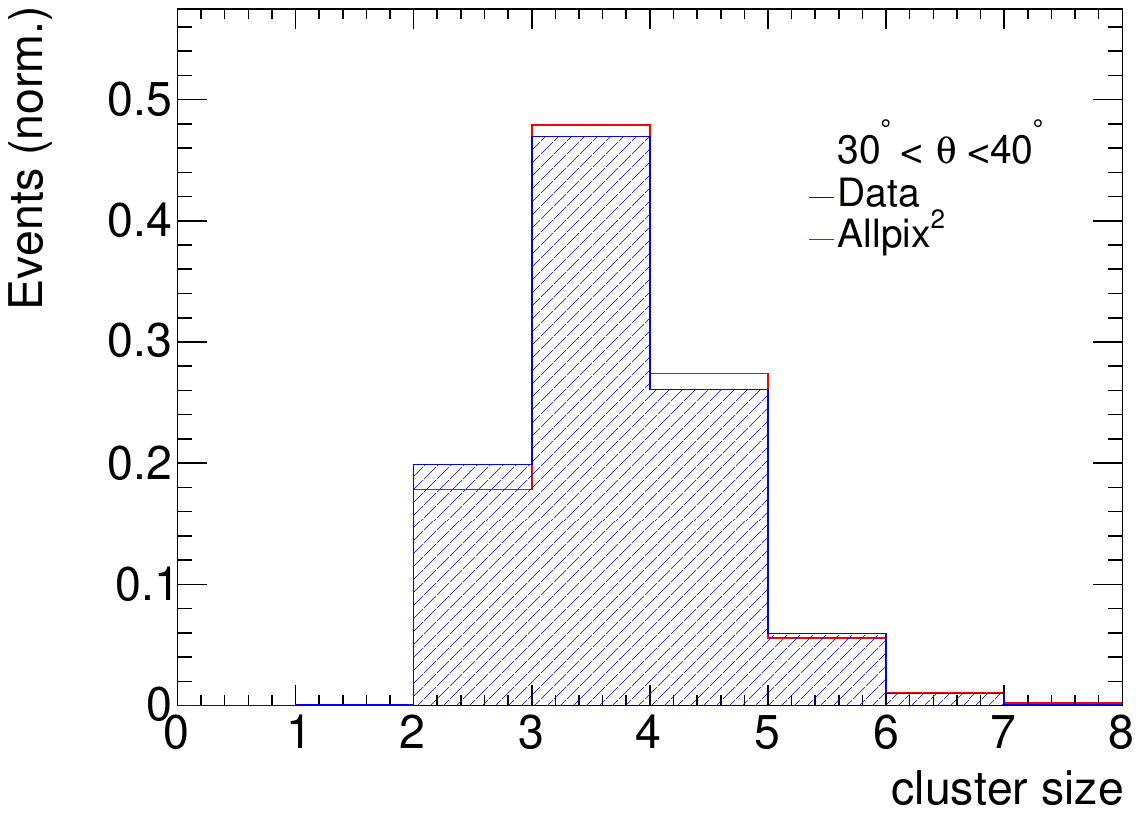}
\caption{The cluster size comparisons for helium nuclei in different $\theta$ bins.}
\label{he_num}
\end{figure}

For validation purposes, the MIP entries from the on-orbit data, consisting of 5.7 million protons and 0.5 million helium, and the simulation data, consisting of 9 million protons and 8 million helium, are selected and compared based on the total ADC count for clusters and cluster size. The distributions are obtained for four intervals of incident angles: $0^{\circ}<\theta<10^{\circ}$, $10^{\circ}<\theta<20^{\circ}$, $20^{\circ}<\theta<30^{\circ}$, and $30^{\circ}<\theta<40^{\circ}$, where $\theta$ represents the projected angle onto a plane defined by the Z-axis and X (Y)-axis. The results for protons are presented in Figures \ref{h_all} to \ref{h_num}, and the results for helium are presented in Figures \ref{he_all} to \ref{he_num}.

\begin{figure}[!htbp]
\centering
\includegraphics[scale=0.34]{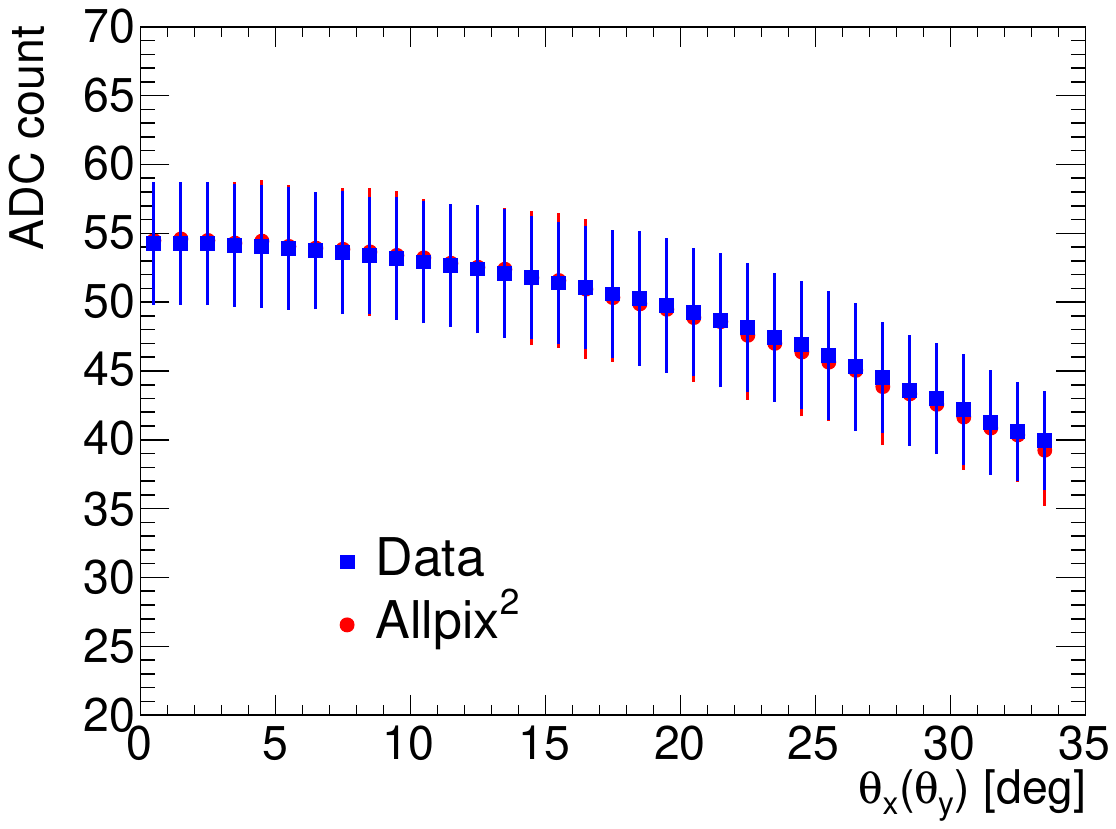}
\includegraphics[scale=0.34]{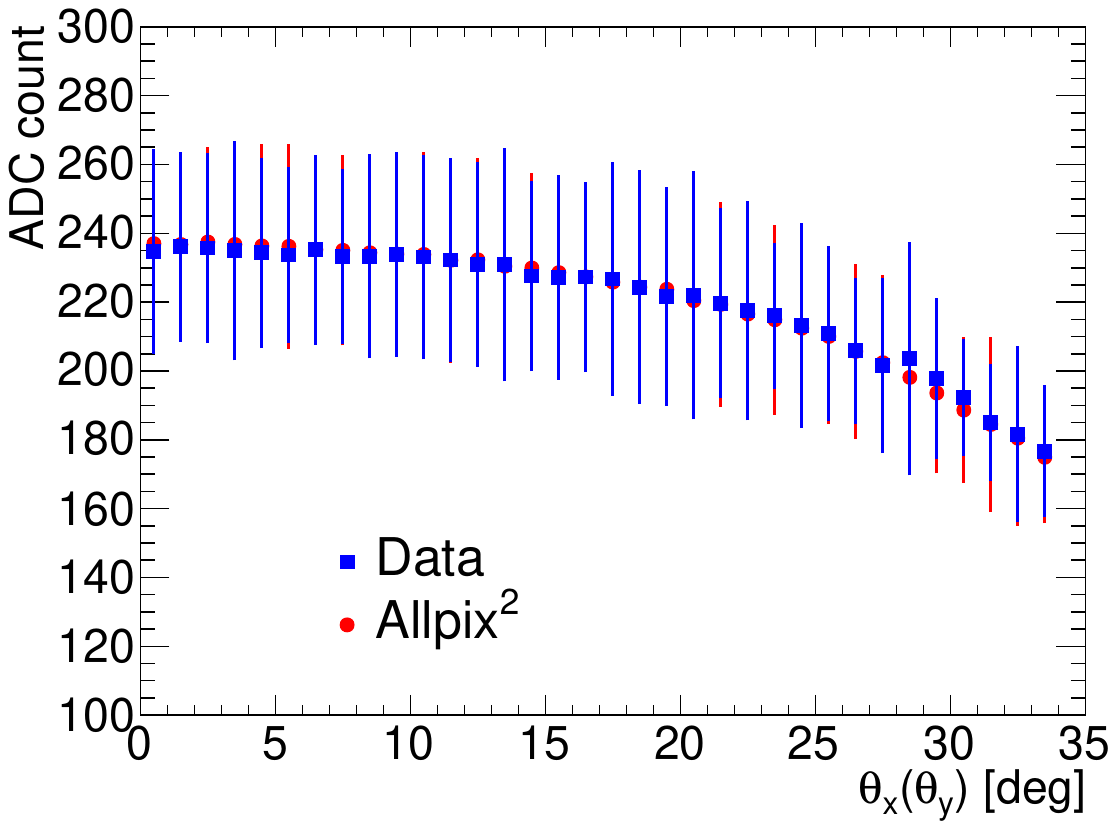}
\caption{The responses of protons (left) and helium (right) hitting a single readout strip at different angles. The error bars in the graph represent the full width at half maximum (FWHM) of the distribution.}
\label{fig:quant}
\end{figure}

In order to quantitatively describe the differences between simulations and on-orbit data, we compared the responses of proton and helium hitting a single readout strip at different angles, as shown in Figure~\ref{fig:quant}.

\section{Conclusions}\label{sec:summary}
The developed simulation tools are valuable for optimizing detector design and operating conditions, as well as for the analysis and interpretation of measurements conducted with silicon strip detectors. In this work, we utilized the $\rm Allpix^{2}$ framework to perform simulations of the STK, and we observed good agreement between the data recorded in beam test and the simulation results. We only utilized a subset of the available modules provided by $\rm Allpix^{2}$.

The $\rm Allpix^{2}$ framework provides a modular approach that allows for wide-ranging applications. By integrating software such as Geant4, TCAD, SPICE, and ROOT, it enables the simulation of the entire process, from particle traversal through the detector to hits in the readout chip. The more we understand the specific parameters of the hardware, the more precise results we can obtain. The reference \cite{allpix_sim} indicates that considering the internal electric field of the silicon strip detector, $\rm Allpix^{2}$ can obtain better simulation results that match the data. This is very helpful for algorithms that require high precision. Moreover, by adjusting different internal parameters of the detector, we can observe variations in overall performance. This aspect holds significant importance for enhancing our understanding of existing detectors and for the development of new detectors.

\section*{Acknowledgements}
This work is supported by the National Key Research and Development Program of China (No. 2022YFF0503301), the National Natural Science Foundation of China (Nos. 12227805, 11873020, 11973097,  12103095, 12235012, U1831206), the Scientific Instrument Developing Project of the Chinese Academy of Sciences (No. GJJSTD20210009), the Youth Innovation Promotion Association of CAS and the Young Elite Scientists Sponsorship Program by CAST (No. YESS20220197).

\bibliography{reference}
\bibliographystyle{elsarticle-num}

\end{document}